\documentclass[twocolumn,10pt]{IEEEtran}

\usepackage{bm,cite,float,amsmath,amssymb,amsthm}
\usepackage{booktabs}
\usepackage[noend]{algpseudocode}
\usepackage{indentfirst}
\usepackage{makecell}
\usepackage{multirow}
\usepackage{graphicx}
\usepackage{subfigure}
\usepackage{multirow}
\usepackage[table]{xcolor}
\usepackage{lipsum}
\usepackage{mathtools}
\usepackage{cuted}
\usepackage[table]{xcolor}
\makeatletter
\def\BState{\State\hskip-\ALG@thistlm}
\makeatother

\usepackage{amssymb}
\usepackage{amsmath}
\usepackage{graphicx}
\usepackage{cite}

\usepackage{balance}
\usepackage[utf8]{inputenc}

\IEEEoverridecommandlockouts

\usepackage{graphicx,epstopdf}
\usepackage{epsfig}	
\usepackage{amsfonts,balance}
\usepackage{bbm}
\usepackage[ruled, linesnumbered]{algorithm2e}
\newcommand{\subparagraph}{}
\usepackage{lipsum}

\newtheorem{theorem}{Theorem}

\newtheorem{corollary}{Corollary}

\makeatletter
\def\ScaleIfNeeded{%
	\ifdim\Gin@nat@width>\linewidth \linewidth \else \Gin@nat@width
	\fi } \makeatother

\makeatletter
\def\underbracex#1#2{\mathop{\vtop{\m@th\ialign{##\crcr
				$\hfil\displaystyle{#2}\hfil$\crcr
				\noalign{\kern3\p@\nointerlineskip}%
				#1\crcr\noalign{\kern3\p@}}}}\limits}

\def\upbracefilla{$\m@th \setbox\z@\hbox{$\braceld$}%
	\bracelu\leaders\vrule \@height\ht\z@ \@depth\z@\hfill 
	\kern\p@\vrule \@width\p@\kern\p@\vrule \@width\p@\kern\p@\vrule \@width\p@
	$}

\def\upbracefillb{$\m@th \setbox\z@\hbox{$\braceld$}%
	\vrule \@width\p@\kern\p@\vrule \@width\p@\kern\p@\vrule \@width\p@\kern\p@
	\leaders\vrule \@height\ht\z@ \@depth\z@\hfill\bracerd
	\braceld\leaders\vrule \@height\ht\z@ \@depth\z@\hfill
	\kern\p@\vrule \@width\p@\kern\p@\vrule \@width\p@\kern\p@\vrule \@width\p@
	$}

\def\upbracefillc{$\m@th \setbox\z@\hbox{$\braceld$}%
	\vrule \@width\p@\kern\p@\vrule \@width\p@\kern\p@\vrule \@width\p@\kern\p@
	\leaders\vrule \@height\ht\z@ \@depth\z@\hfill
	\kern\p@\vrule \@width\p@\kern\p@\vrule \@width\p@\kern\p@\vrule \@width\p@
	$}

\def\upbracefilld{$\m@th \setbox\z@\hbox{$\braceld$}%
	\vrule \@width\p@\kern\p@\vrule \@width\p@\kern\p@\vrule \@width\p@\kern\p@
	\leaders\vrule \@height\ht\z@ \@depth\z@\hfill\braceru$}

\def\upbracefillbd{$\m@th \setbox\z@\hbox{$\braceld$}%
	\vrule \@width\p@\kern\p@\vrule \@width\p@\kern\p@\vrule \@width\p@\kern\p@
	\bracerd\braceld
	\leaders\vrule \@height\ht\z@ \@depth\z@\hfill\braceru$}

\makeatother
\begin{document}
	%Heterogeneous Federated Reinforcement Learning for Uplink Centric Broadband Communication Optimization over Unlicensed Spectrum
	%DRL-based Channel Access Optimization over Unlicensed Spectrum for Uplink Centric Broadband Communication
	%DRL-based Unlicensed Channel Access Optimization for Uplink Centric Broadband Communication
	%Federated Reinforcement Learning for Uplink Centric Broadband Communication Optimization over Unlicensed Spectrum
	%DRL-based Channel Access Optimization over Unlicensed Spectrum for UCBC
	\title{Federated Reinforcement Learning for Uplink Centric Broadband Communication Optimization over Unlicensed Spectrum
		%		\thanks{This work was supported by the King’s Together Multi and Interdisciplinary Research Scheme.}
	}
	
	\author{Hui~Zhou,
		Yansha~Deng,~\IEEEmembership{Senior Member,~IEEE.}
		
		\vspace{-0.2cm}
		\thanks{
			H. Zhou and Y. Deng are with  Department of Engineering, King's College London, London, WC2R 2LS, UK (email:\{hui.zhou, yansha.deng\}@kcl.ac.uk)(Corresponding author: Yansha Deng).}
		
		%\thanks{K. C. Cheung is with Department of Electrical and Computer Engineering,
		%University of British Columbia, Vancouver, BC, , V6T 1Z4, Canada (email: \{kcheung\}@ece.ubc.ca).}
		%\thanks{ R. Schober is with the Department of Electrical and Computer Engineering,
		%University of British Columbia, Vancouver, BC, Canada, V6T 1Z4, and also
		%with  the Institute for Digital Communication, Friedrich-Alexander-Universit\"{a}t Erlangen-N\"{u}rnberg (FAU), Erlangen, Germany (email: \{rschober\}@ece.ubc.ca; \{schober\}@lnt.de).}
	}
	
	\maketitle
	
	\begin{abstract}
		To provide Uplink Centric Broadband Communication (UCBC), New Radio Unlicensed (NR-U) network has been standardized to exploit the unlicensed spectrum using Listen Before Talk (LBT) scheme to fairly coexist with the incumbent Wireless Fidelity (WiFi) network. Existing access schemes over unlicensed spectrum are required to perform Clear Channel Assessment (CCA) before transmissions, where fixed Energy Detection (ED) thresholds are adopted to identify the channel as idle or busy. However, fixed ED thresholds setting prevents devices from accessing the channel effectively and efficiently, which leads to the hidden
		node (HN) and exposed node (EN) problems. In this paper, we first develop a centralized double Deep Q-Network (DDQN) algorithm to optimize the uplink system throughput, where the agent is deployed at the central server to dynamically adjust the ED thresholds for NR-U and WiFi networks. Considering that heterogeneous NR-U and WiFi networks, in practice, cannot share the raw data with the central server directly, we then develop a federated DDQN algorithm, where two agents are deployed in the NR-U and WiFi networks, respectively. Our results have shown that the uplink system throughput increases by over 100\%, where cell throughput of NR-U network rises by 150\%, and cell throughput of WiFi network decreases by 30\%. To guarantee the cell throughput of WiFi network, we redesign the reward function to punish the agent when the cell throughput of WiFi network is below the threshold, and our revised design can still provide over 50\% uplink system throughput gain.	
%		Finally, we validate the effectiveness of proposed DRL algorithm under the coexistence of WiFi networks.
		
	\end{abstract}
	
	
	\begin{IEEEkeywords}
		Federated learning, deep reinforcement learning, NR-U, UCBC, energy detection.
	\end{IEEEkeywords}
	
	\section{Introduction}

	With the popularity of emerging data-hungry applications, including high-definition (HD) video surveillance, and unmanned aerial vehicle (UAV) \cite{Hu2021, Zhou2021}, Uplink Centric Broadband Communication (UCBC) has been identified as new service class in the vision of 5.5G to cater for the exponentially growing uplink traffic volume, which leads to a tremendously heavy burden on the existing cellular network due to scarce licensed spectrum and downlink traffic-oriented design \cite{Letaief2022,Huawei}. To cope with the ever-increasing uplink traffic demand, the Third Generation Partnership Project (3GPP) has proposed to boost the uplink capacity by exploiting the unlicensed spectrum. In particular, the 3GPP has standardized Long Term Evolution (LTE) based technologies including LTE licensed-assisted access (LTE-LAA) \cite{Mukherjee2016,Kwon2017} and enhanced LAA (eLAA) \cite{Chen2019} to access the unlicensed spectrum since Release 13. Recently, New Radio Unlicensed (NR-U) network has been standardized to extend the applicability of New Radio (NR) to unlicensed spectrum in Release 16, which determines to support standalone operation mode over unlicensed spectrum \cite{Hirzallah2021,Lagen2020,Naik2020}.
	
	%tremendously push the use of spectrum resource to the limit. scarce bandwidth of licensed spectrum .,  The Fifth
	
	Although the unlicensed spectrum is a promising solution to provide the cellular network with sufficient uplink spectrum resources, no exclusive rights are granted to any networks in the unlicensed spectrum, which means the NR-U network needs to coexist with the incumbent Wireless Fidelity (WiFi) network \cite{Bianchi2000}. To fairly and harmoniously coexist with the WiFi network, the NR-U network adopts the contention-based access scheme, called Listen Before Talk (LBT), to access the unlicensed spectrum, where the LBT scheme incorporates the Backoff procedure similar to the Carrier-sense multiple access with collision avoidance (CSMA/CA) scheme in WiFi network.

	%proposing new access schemes \cite{Campos2018,KSheshadri2018,Lagen2018,Li2017} orFor newly proposed access schemes, the short-term Channel Quality Indicator (CQI) has been identified as the potential tool to detect the presence of HN problem in multiple LTE-LAA scenario, which is expected to have an oscillating behavior \cite{Campos2018}. In \cite{KSheshadri2018}, pilot and data transmission have been utilized to detect the HN problem in LTE-LAA, where the devices are required to periodically send control packets. In \cite{Lagen2018}, a paired LBT solution has been proposed to improve the throughput while avoiding the HN problem in unlicensed millimeter-wave (mmWave) bands. In \cite{Li2017}, the authors have proposed to adaptively change the ED threshold of LTE-LAA, where the ED threshold is decreased by one everytime the transmission fails. However, all these newly proposed schemes have not been incorporated into the 3GPP standard and verified with adequate experiments. For the access control parameters optimization,
	
	With the goal of improving the throughput over the unlicensed spectrum, the majority of efforts in previous works have been devoted to explicitly optimizing access control parameters using the formulated mathematical model \cite{Tan2019,Bao2021,Liu2019}. The authors in \cite{Tan2019} have optimized the transmission time of the LBT scheme with fixed contention window size to maximize the number of devices with guaranteed Quality of Service (QoS), where the combinatorial and nonconvex problem is decomposed into two sub-problems to obtain the sub-optimal solution. In \cite{Bao2021}, the iterative algorithm has been proposed to optimize the time and power allocation, where devices are assumed to be able to detect each other without HN problem. In \cite{Liu2019}, the authors have proposed a hybrid duty-cycle and LBT access scheme over the unlicensed spectrum, where the contention window and duty-cycle time fraction are optimized via exhaustive searching method and iterative algorithm, respectively. In \cite{Peng2022, Daraseliya2021}, the authors derived the optimal initial optimal Backoff window size, where Markov chain was utilized to obtain the successful transmission propobility in steady state. However, due to the mathematical complexity of the formulated non-convex problem, all these works optimized the access control parameters with the sub-optimal solution under the impractical assumption for simplicity. 
%	In \cite{Daraseliya2021}, the authors constructed an mathematical framework to characterize the achievable data rates over unlicensed spectrum. where the optimal initial contention window size is derived.

	To deal with more complex communication environment and practical formulations, Machine Learning (ML), especially Reinforcement Learning (RL) \cite{jiang2021a}, emerges as a promising tool to optimize access control parameters over the unlicensed spectrum, due to that it solely relies on the self-learning of the environment interaction, without the need to derive explicit optimization solutions based on a complex mathematical model. In \cite{Zhang2020}, each device can be controlled to transmit or remain silent in every slot based on Deep Reinforcement Learning (DRL) and Federated Learning (FL) approaches, where no NR-U network and HN problem are considered for simplicity. In \cite{Challita2018}, the authors have proposed to utilize the DRL to optimize the airtime fraction for LTE-LAA without considering the detailed LBT procedures. In \cite{Tan2020}, a duty-cycle free spectrum sharing framework has been designed, where the DRL algorithm is utilized to maximize the throughput via optimizing the LTE transmission time. In \cite{Tan2018}, the transmission time of frame-based LBT has been optimized via RL, where each device independently chooses the access technologies, including LTE, LTE-LAA, and WiFi. In \cite{Ali2021}, RL is combined with FL to optimize the contention window size of WiFi network with the aim to minimize the collision probability. In \cite{Ali2017}, the authors have proposed to adjust the contention window size with a minimum delay, where the neural network is utilized to predict the number of Negative Acknowledges (NACKs). In \cite{Pei2022}, the authors proposed a multi-agent RL learning algorithm to optimize the access time and transmission time among Base Stations (BSs), where the mean field technology was utilized to solve the Nash Equilibrium (NE).
	
	%Knowing the ED thresholds to be the key point of HN and EN problems, it is important to dynamically adjust the ED thresholds based on the real-time wireless environment, which has the potential to boost the throughput via mitigating the HN and EN problems \cite{Iqbal2017}. 

		However, the above learning-based access control solutions either focused on optimizing the backoff random number or the contention window size in the downlink transmission, without considering the Uplink Centric Broadband Communication (UCBC) over unlicensed spectrum. It is noted that optimizing the contention window size and backoff number fails to solve the hidden node and exposed node problems in the uplink transmission due to the following two reasons: 1) optimizing the contention window and backoff random number only changes the access time of NR-U UEs and WiFi STAs when the collision happens. The devices with fixed energy detection threshold will continue to identify the channel as idle and transmit data when the channel is occupied by other devices for a long time, which may still lead to collision; 2) each UE maintains a different contention window size and backoff number independently in the uplink transmission, which means optimizing the contention window size and backoff number in the uplink transmission requires much higher number of agents (i.e., proportional to the number of the UEs ) than the downlink (i.e., proportional to the number of BSs). Therefore, we propose federated reinforcement learning algorithm to dynamically optimize the common energy detection threshold adopted by all UEs.  Furthermore, the existing works merely evaluated system throughput to indicate the network-centric performance, and ignored the user-centric Quality of Experience (QoE)\cite{Bairagi2018, Bairagi2019}, which is more intuitive to reveal individual user experience in the UCBC.
	
	In this article, we address the following fundamental questions: 1) how to optimize the uplink system throughput over unlicensed spectrum under the coexistence of heterogeneous networks; 2) how to guarantee the fairness between heterogeneous networks during optimization; 3) how to evaluate the effectiveness of proposed algorithm in improving the user-centric QoE performance. To do so, we develop DRL-based ED thresholds configuration approaches to dynamically optimize the uplink system throughput under the coexistence of NR-U and WiFi networks. Our contributions can be summarized as follows:
	\begin{itemize}
		\item We develop a DRL-based framework to optimize the uplink system throughput by adaptively configuring the ED thresholds under the coexistence of heterogeneous NR-U and WiFi networks. In the framework, the uplink transmission procedure over the unlicensed spectrum is simulated by taking into account the traffic characteristic of UCBC, the process of LBT and CSMA/CA access schemes, uplink transmission scheduling in NR-U network, and the collision among devices, which is used for training the DRL-based agent before deployment.
		%According to the 3GPP standard, the gNB performs the Cat4 LBT to schedule the UE, then the UE can either perform Cat2 LBT or Cat4 to transmit the packet, which are termed as gNB-Cat4/UE-Cat2 scheme and gNB-Cat4/UE-Cat4 scheme, respectively,
		\item We first propose a centralized double Deep Q-Network (DDQN) algorithm to optimize the uplink system throughput via dynamically configuring the ED thresholds for both the NR-U and WiFi networks, where the agent is deployed at the central server to observe the number of successfully transmitted packets of both networks during training and testing phase in real-time. Two different NR-U uplink transmission procedures are considered according to the 3GPP standard \cite{NRU2018}, where the UE can either perform Category (Cat) 2 LBT or Cat4 LBT to transmit the packet. 
		\item To protect the data privacy of the NR-U and WiFi networks, we then develop a federated DDQN algorithm via Federated Averaging (FedAve), where two agents are deployed at the NR-U and WiFi networks, respectively. Two agents exchange the model parameters with the central server, and can be independently deployed in NR-U and WiFi networks during the testing phase \cite{McMahan2017}. It is noted that we introduce the user-centric performance metric, user perceived throughput (UPT), to evaluate the file transmission throughput improvement from the perspective of users.
		\item As the NR-U network has inherent advantages over the WiFi network, including higher data rate and scheduling procedure, the agent tends to sacrifice the cell throughput of WiFi network for higher uplink system throughput under the coexistence scenario. In order to take the fairness into account, we redesign the reward function to guarantee the cell throughput of WiFi network, where the agent gets punished when the cell throughput of WiFi network is lower than the pre-defined threshold. 
		\item Finally, our proposed DRL-based learning framework is evaluated to optimize the uplink system throughput in heterogeneous NR-U and WiFi networks. Without considering the fairness, the results have shown that the system throughput achieves up to 100\% performance gain, where the cell throughput of NR-U network increases by 150\%, and the cell throughput of WiFi network decreases by 30\%. Taking the fairness into account, our results,  the results have shown that the cell throughput of WiFi network is even higher than the benchmark scheme with fixed ED thresholds, where the system throughput achieves 70\% performance gain under File Transfer Protocol 3 (FTP-3) traffic, and 50\% performance gain under more dynamic Beta traffic. In terms of the UPT, the performance gain of NR-U network achieves over 100\% in all experimental settings.
		
%		To further verify the effectiveness of the proposed algorithm, we also consider the coexistence of WiFi networks, where the proposed algorithm still performs better than the benchmark scheme with fixed ED thresholds.
	\end{itemize}
	
	The remainder of this paper is organized as follows. Section II provides the system model of the NR-U and WiFi networks. Section III describes the problem analysis and formulation. Section IV elaborates the proposed centralized DRL and federated DRL algorithms for solving the formulated problem. Section V illustrates the numerical results. Finally, Section VI concludes the paper.
	
	\section{System Model}
\begin{figure}[!h]
		\centerline{\includegraphics[scale=0.5]{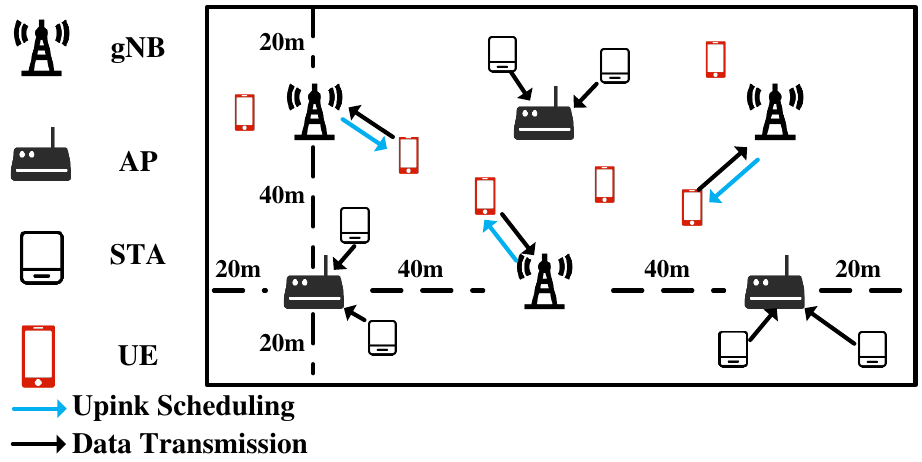}}
		\caption{Uplink transmission of NR-U and WiFi indoor coexistence scenario.}
		\label{Scenario}
	\end{figure}

	We consider the uplink transmission of the 3GPP indoor scenario as shown in Fig.~\ref{Scenario}, where NR-U and WiFi networks deploy three small cells in a one-floor building, respectively\cite{NRU2018}. It is noted that the heterogeneous NR-U and WiFi networks share a single 20-MHz unlicensed channel for transmission. The set of WiFi stations (STAs) and NR-U User Equipments (UEs) are denoted by $\mathcal{W}$ and $\mathcal{U}$, respectively. The set of access points (APs) and gNodeBs (gNBs) are denoted by $\mathcal{P} $ and $\mathcal{G}$, respectively. We assume that STAs  and UEs are uniformly distributed in the scenario, where each STA or UE is connected to the closest AP or gNB. The main notations are summarized in Table ~\ref{Notations}.
	
	\begin{table}[h!]
		\caption{Notation Table}
		\begin{center}
			\begin{tabular}{|c|c|}
				\hline 
				\rowcolor{gray}\textbf{Notations} & \textbf{Physical means} \\
				\hline
				$\mathrm{S_\mathrm{file}}$ & File Size \\
				\hline
				$\mathrm{T_{con}}$ & Time Constraint \\
				\hline
				$\mathrm{CW_{max}}$ & Maximum Contention Window\\
				\hline
				${\mathrm{CW_{min}}}$&Minimum Contention Window\\
				\hline
				$N$&Backoff Number\\
				\hline
				$\lambda$ & Packet Arrival Rate \\
				\hline
				$\mathrm{P}_{\mathrm{\{S,U\}}}$&Transmit Power of UE and STA \\
				\hline
				$\mathrm{P}_{\mathrm{\{G\}}}$&Transmit Power of gNB \\
				\hline
				$\sigma_{n}^2$&Noise Power \\
				\hline
				$h$& Channel Power Gain\\
				\hline
				$\alpha$&Transmission Indicator\\
				\hline
				$\mathcal{I}$&Interference\\
				\hline
				$\eta_{\mathrm{w}}$&WiFi SINR Threshold \\
				\hline
				$\eta_{\mathrm{n}}$&NRU SINR Threshold \\
				\hline
				$\lambda^{\mathrm{CS}}_{\mathrm{w}}$&WiFi Preamble Detection Threshold \\
				\hline
				$\lambda^{\mathrm{ED}}_{\mathrm{w}}$&WiFi Energy Detection Threshold  \\
				\hline
				$\lambda^{\mathrm{ED}}_{\mathrm{n}}$&NRU Energy Detection Threshold \\
				\hline
			\end{tabular}
			\label{Notations}
		\end{center}
	\end{table}
	
	\subsection{Network and Traffic Model}
	We consider a flat Rayleigh small-scale fading channel, where the channel power gains $\beta(x,y)$ between two generic locations $ x, y \in \mathbb{R}^3 $ is assumed
	to be exponentially distributed random variables with unit mean. All the channel gains are independent of each other, independent of the spatial locations, and identically distributed (i.i.d.). 
	
	The indoor mixed office path-loss model is adopted as \cite{channel_model_2019}
	\begin{equation}
	\begin{aligned}
	\zeta_\mathrm{L}(x,y)=
	32.4+17.3\log_{10}(d_{\mathrm{3D}})+20\log_{10}(f_\mathrm{c}),
	\end{aligned}
	\label{eq:los}
	\end{equation}
	\begin{equation}
	\begin{aligned}
	\zeta_\mathrm{N}(x,y)=
	32.4+31.9\log_{10}(d_{\mathrm{3D}})+20\log_{10}(f_\mathrm{c}),
	\end{aligned}
	\label{eq:nlos}
	\end{equation}
	where $\zeta_\mathrm{L}(x,y)$ and $\zeta_\mathrm{N}(x,y)$ represent the pathloss under line-of-sight (LoS) and non line-of-sight (NLoS), $f_\mathrm{c}$ is the carrier frequency, and $d_\mathrm{3D}$ is the distance between two locations $x$ and $y$. The indoor mixed office LoS probability $P_\mathrm{LoS}$ is given as
	\begin{equation}
	P_\mathrm{LoS}=\begin{cases}
	1&,d_\mathrm{2D} \textless 1.2\mathrm{m}\\
	\exp(\dfrac{d_\mathrm{2D}-1.2}{4.7})& ,1.2 \mathrm{m} \leq d_\mathrm{2D} \leq 6.5 \mathrm{m}\\
	\exp(\dfrac{d_\mathrm{2D}-6.5}{32.6})& ,d_\mathrm{2D} \textgreater 6.5 \mathrm{m}
	\end{cases}
	\end{equation}
	where $d_\mathrm{2D}$ is the projection of $d_\mathrm{3D}$ on the horizontal plane. Accordingly, the NLoS probability $P_\mathrm{NLoS}$ is
	\begin{equation}
	P_\mathrm{NLoS}=1-P_\mathrm{LoS}.
	\end{equation}
	
	Therefore, the mean channel power gain is derived as
	\begin{equation}
	h=(10^{-\zeta_\mathrm{L}/10}P_\mathrm{LoS} + 10^{-\zeta_\mathrm{N}/10}P_\mathrm{NLoS})\beta,
	\end{equation}
	where the spatial indices $ (x, y) $ are dropped for the brevity of exposition.
	
	To capture the burst traffic characteristic of UCBC, we consider the FTP-3 traffic model for each STA and UE with fixed size $\mathrm{S_\mathrm{file}}$, where packets arrive according to a Poisson process with arrival rate $\lambda$\cite{LAA2015,TrafficFTP32017}. Without loss of generality, First Come First Serve (FCFS) scheduling scheme is applied by placing the newly arrived packets at the end of the queue. We assume each packet has a time constraint $\mathrm{T_{con}}$, where the packet is dropped if it is not successfully transmitted within the time constraint.
	
	%\begin{figure*}[!h]
	%	\centerline{\includegraphics[scale=0.9]{Timing graph of NR-U and WiFi.pdf}}
	%	\caption{Timing graph of access procedure for WiFi and NR-U.}
	%	\label{Timing_graph}
	%\end{figure*}

	\begin{figure*}[!h]
		\centering
		\subfigure[Uplink transmission of WiFi network with CSMA/CA scheme]{\includegraphics[width=14cm]{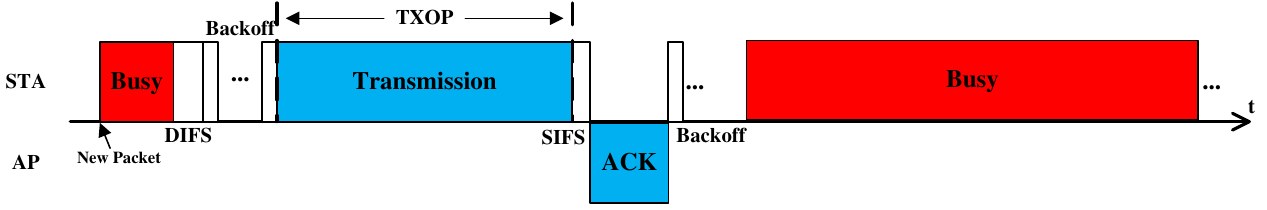}} 
		\\
		\centering
		\subfigure[Uplink transmission of NR-U network with UE Cat4 LBT scheme]{\includegraphics[width=14cm]{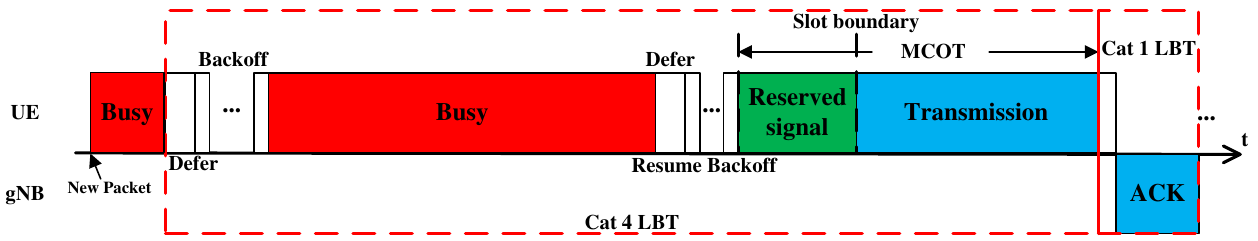}}
		\caption{Timing graph of uplink transmission for WiFi and NR-U networks.}
		\label{Timing_graph}
	\end{figure*}
	
	\subsection{Channel Access Schemes in Unlicensed Band}
	As heterogeneous WiFi and NR-U networks coexist over the unlicensed spectrum with different access schemes, we present the detailed procedures of the CSMA/CA and LBT access schemes in this section, respectively.
	\subsubsection{Carrier-Sense Multiple Access with Collision Avoidance}
	The WiFi network adopts the CSMA/CA to access the unlicensed spectrum \cite{Ho2017}, also known as distributed coordination function (DCF), which integrates the Backoff procedure as shown in Fig.~\ref{Timing_graph}~(a). When the new packet arrives, the STA is required to perform CCA to determine whether the channel is busy or idle. If the channel has been continuously idle over a DCF inter-frame space (DIFS) interval, it transmits the packet immediately. Otherwise, the STA defers its transmission until the channel becomes idle. Then if the channel is detected to be continuously idle over a DIFS interval, the STA will initiate the Backoff procedure to further defer its transmission over a random time interval. The back-off procedure starts with the selection of an integer $N$, where $N$ is a random number uniformly distributed in the range from 0 to the contention window $\mathrm{CW}$. It is noted that the $\mathrm{CW}$ is initialized to be the minimum value ${\mathrm{CW_{min}}}$. Next, the generated random number $N$ decreases when the CCA identifies the channel to be idle. Otherwise, the random number $N$ is frozen, and continues to decrease when the CCA identifies the channel to be idle again within a DIFS interval. Once $N$ reaches zero, the STA transmits its data within the transmission opportunity (TXOP). It is important to note that this Backoff procedure randomizes the channel access among STAs and thus helps to reduce the chance of collision.
	
	Upon receiving the packet correctly, the AP waits for a Short inter-frame space (SIFS) interval, and transmits an ACK back to the STA to confirm the correct reception. When the transmission is successful, the contention window $\mathrm{CW}$ is reset to its minimum value ${\mathrm{CW_{min}}}$. Otherwise, the STA activates the retransmission procedure for the lost packet, where the contention window size $\mathrm{CW}$ is doubled until it reaches a maximum value ${\mathrm{CW_{max}}}$. 
	
	\subsubsection{Listen Before Talk}
	To fairly share the unlicensed spectrum with the incumbent WiFi network, the NR-U network has agreed to adopt the LBT schemes, which consist of following three categories:
	\begin{itemize}
		\item Cat4: LBT with random Backoff and a contention window of variable size,
		\item Cat2: LBT without random Backoff,
		\item Cat1: Immediate transmission after a short switching gap.
	\end{itemize}
	
	As shown in Fig.~\ref{Timing_graph}~(b), the UE adopts the Cat4 LBT channel access scheme to transmit the user data using the scheduled Physical Uplink Shared Channel (PUSCH) resource. Inheriting the Backoff procedure from CSMA/CA, the UE is required to perform CCA to check whether the channel has been continuously idle over a Defer interval when the new packet arrives. If the channel has been idle for a Defer interval, it transmits immediately. Otherwise, the UE defers its transmission until the channel becomes idle. Once the channel is detected to be continuously idle over a Defer interval, the UE will initiate the Backoff procedure to further defer its transmission over a random time interval as CSMA/CA. Once the random number $N$ finally reaches zero, the UE can transmit its data within the maximum channel occupancy time (MCOT). It is noted that the UE can only start data transmission at the spectrum slot boundary (SSB). Hence, if the Backoff procedure does not end at the SSB, the UE is required to transmit a reserved signal (RS) until the SSB.
	
	Although the Backoff procedure reduces the collision probability, it also leads to additional controlling overheads. Therefore, 3GPP determines to support the Cat2 LBT channel access scheme, which only needs to guarantee the channel to be idle for 25$us$ continuously before transmission without Backoff procedure. 
	
	The Cat1 LBT channel access scheme is referred to as ``no LBT". More specifically, if the gap between two successive transmissions is less than or equal to 16$us$, the latter transmission does not need to perform any LBT. As shown in Fig.~\ref{Timing_graph}~(b), Cat 1 LBT is beneficial for NR-U to support fast ACK/NACK feedback transmission, where the gap between PUSCH and Physical Downlink Control Channel (PDCCH) is less or equal to 16$us$.

	\subsection{Key Differences between NR-U and WiFi Networks}
	Although the Cat4 LBT access schemes adopt similar procedures as CSMA/CA to guarantee the fairness, there are fundamental differences between the WiFi and NR-U networks in terms of the frame structure, uplink scheduling, and access category, which are presented in detail in this section.
	\subsubsection{Frame Structures}
	As we mentioned above, the WiFi STA can start the data transmission once the random number $N$ decreases to zero. However, the NR-U UE can only start the data transmission at the exact SSB, which can hardly be guaranteed due to the random nature of Backoff procedure\cite{Loginov2021}. Currently, 3GPP has not explicitly defined the behaviour between the end of Backoff and SSB. Considering that if the NR-U UE decides to wait for the SSB without transmission, WiFi STAs may identify the channel as idle and occupy the channel, which leads to poor NR-U network performance. Therefore, it is usually suggested to send an RS to occupy the channel until the SSB \cite{Kim2020}. 
	
	However, the transmission of RS introduces the controlling overheads, especially with larger slot length (e.g., with a period $\theta= 500us$ in the LTE-LAA network). It is noted that the problem can be alleviated in NR-U network based on its flexible radio numerology, where the NR-U network adopts higher 
	subcarrier spacing (SCS), and even mini-slot to reduce the slot length and RS overheads \cite{KosekSzott2021}.
	
	\subsubsection{Uplink Transmission Procedure}
	In the WiFi network, each STA performs the CSMA/CA to access the unlicensed channel once the new packets arrive, where the AP does not schedule the STA to transmit. Each WiFi STA accesses the channel independently, and transmits the uplink user data once it obtains the channel.  However, in each NR-U cell, NR-U UEs are scheduled for uplink transmission in an centralized way, where each UE needs to be scheduled individually before transmitting the user data using the granted PUSCH resource. Therefore, there is only inter-cell interference between the NR-U UEs without intra-cell interference.
	
\begin{figure}[!h]
		\centering
		\subfigure[Uplink transmission of NR-U network with gNB-Cat4/UE-Cat2 LBT]
		{\includegraphics[scale=0.65]{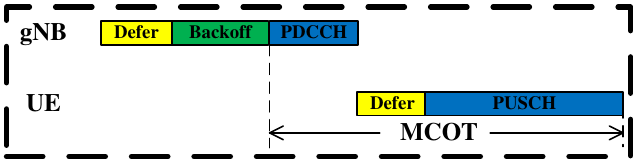}} 
		\\
		\centering
		\subfigure[Uplink transmission of NR-U network with gNB-Cat4/UE-Cat4 LBT]
		{\includegraphics[scale=0.5]{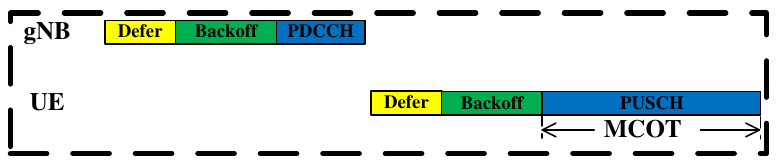}}
		\caption{Procedures of gNB scheduling and UE data transmission.}
		\label{Uplink_shceduling}
	\end{figure}

	The 3GPP standard has identified uplink transmission procedure as shown in the Fig.~\ref{Uplink_shceduling}, where the gNB is required to perform the Cat4 LBT to schedule the UE via PDCCH. Then the UE can either perform Cat2 LBT as Fig.~\ref{Uplink_shceduling}(a) or Cat4 LBT as Fig.~\ref{Uplink_shceduling}(b) to transmit the user data on the granted PUSCH resource \cite{NRU2018}. For simplicity, these two uplink transmission schemes are termed as gNB-Cat4/UE-Cat2 and gNB-Cat4/UE-Cat4, respectively. It is noted that when the UE performs the Cat2 LBT, the UE can access the channel after Defer time without performing Backoff procedure, which can be adopted for latency-sensitive service to guarantee the stringent latency requirement.
	
	\subsubsection{Access Category}
	As the basic DCF with CSMA/CA scheme lacks capabilities to guarantee the QoS of different applications, the enhanced distribution coordination function (EDCF) has introduced the concept of access category (AC), where four priority levels are defined to differentiate the channel access probability for different traffic types\cite{Zheng2020}. Correspondingly, the LBT access schemes also define four priority levels with different parameters, where shorter wait time and contention window size correspond to higher priority service \cite{Ma2018}.

	\section{Problem analysis and formulation}
	To capture the characteristics of uplink transmission with CSMA/CA and LBT access schemes over unlicensed spectrum, we consider the Carrier Sensing (CS) \& ED, and Signal to Interference Plus Noise Ratio (SINR) to model the CCA and data transmission, respectively, which are then utilized to formulate the system throughput optimization problem under the WiFi and NR-U coexistence scenario. We also introduce a performance metric called UPT to evaluate the user-centric QoE.
	
	\subsection{Carrier Sensing \& Energy Detection}
	In WiFi network, each STA is required to perform CCA during the DIFS interval and Backoff procedure to determine whether the channel is busy or idle in each slot. The CCA in WiFi network consists of the CS \& ED, where the CS detects the preamble transmission of WiFi network, and ED detects the total energy on the channel including both transmissions from NR-U and WiFi networks. We formulate the CS and ED of WiFi STA $i$ $(i \in \mathcal{W})$ as
	\begin{equation}
	{{\mathrm{WiFi}}_{i}^{\mathrm{CS}}} = \sum_{w \in \mathcal{W} \backslash i}{\alpha_{w}\mathrm{P}_{\mathrm{w}}h_{w,i}}
	\label{eq:wifi preamble decoding},
	\end{equation}
	\begin{equation}
	{{\mathrm{WiFi}}_{i}^{\mathrm{ED}}} = \sum_{w \in \mathcal{W} \backslash i}{\alpha_{w}\mathrm{P}_{\mathrm{w}}h_{w,i}}+\sum_{u \in \mathcal{U}}{\alpha_{u}\mathrm{P}_{\mathrm{u}}h_{u,i}}+\sum_{g \in \mathcal{G} }{\alpha_{g}\mathrm{P}_{\mathrm{g}}h_{g,i}}
	\label{eq:wifi energy detection},
	\end{equation}
	where $w \in \mathcal{W}$ denotes the STA, $u \in \mathcal{U}$ denotes the UE, $g \in \mathcal{G}$ denotes the gNB. In \eqref{eq:wifi preamble decoding} and \eqref{eq:wifi energy detection}, $\alpha_{\{w,u,g\}}$ indicates whether device $\{w,u,g\}$ is transmitting or not, $h_{\{w,u,g\},i}$ is the channel gain between device $\{w,u,g\}$ and STA $i$, and $\mathrm{P}_{\{w,u,g\}}$ is the transmit power.
	
	As shown in \eqref{eq:wifi preamble decoding} and \eqref{eq:wifi energy detection}, the STA only senses the transmission of other STAs in the CS, but detects the transmission energy of all other devices in the ED. The STA will only identify the channel as idle when the values of CS and ED are under the pre-defined threshold $\lambda^{\mathrm{CS}}_{\mathrm{w}}=-82 \mathrm{dBm}$ and $\lambda^{\mathrm{ED}}_{\mathrm{w}}=-62 \mathrm{dBm}$, respectively.
	
	In the NR-U network, the NR-U device $i \in \mathcal{U} \cup \mathcal{G}$ is required to perform CCA during the Defer interval and Backoff procedure to determine whether the channel is busy or idle in each slot, which is formulated as
	\begin{equation}
	{{\mathrm{NR}}_{i}^{\mathrm{ED}}} = \sum_{w \in \mathcal{W}}{\alpha_{w}\mathrm{P}_{\mathrm{w}}h_{w,i}}+\sum_{u \in \mathcal{U}\backslash i}{\alpha_{u}\mathrm{P}_{\mathrm{u}}h_{u,i}}+\sum_{g \in \mathcal{G}\backslash i }{\alpha_{g}\mathrm{P}_{\mathrm{g}}h_{g,i}}
	\label{eq:nr energy detection}.
	\end{equation}
	When the ED is below the predefined threshold $\lambda^{\mathrm{ED}}_{\mathrm{n}}=-72\mathrm{dBm}$, the channel is identified as idle.
	
	\subsection{Signal to Interference Plus Noise Ratio}
	The device starts the transmission once finishing the Backoff in WiFi network or at the exact SSB in NR-U network. However, due to the interference from the hidden nodes or channel fading, the receiver may fail to decode the transmitted signal. We model the decoding process via the SINR, where the SINR of data transmission from device $i \in \mathcal{K}$ to the device $j \in \mathcal{K}$ can be represented as
	\begin{equation}
	{{\mathrm{SINR}}_{i,j}} = \frac{\mathrm{P}_{i} h_{i,j}}{\sum_{k \in \mathcal{K} \backslash i}{\alpha_{k,i}\mathrm{P}_{k}h_{k,j}}+\sigma_{n}^2}
	\label{eq:SINR},
	\end{equation}
	where $\mathcal{K}=\mathcal{W} \cup \mathcal{U} \cup \mathcal{G} \cup \mathcal{P}$ represents the set of all devices, $\alpha_{k}$ indicates whether device $k$ is transmitting or not, $h_{i,j}$ is the channel from device $i$ to the device $j$, $\mathrm{P}_{i}$ represents the transmit power of device $i$, and $\sigma_{n}^2$ is the power of the noise.
	
	\subsection{User Perceived Throughput}
	Since the system throughput can only reflect the network-level performance, we introduce a user-centric performance metric called UPT to evaluate the file transmission throughput from the perspective of users. The UPT is obtained by averaging all file throughputs \cite{TrafficFTP32017}, where each file throughput $T_\mathrm{file}$ is calculated as 
		\begin{equation}
	T_\mathrm{file}=\begin{cases}
	\dfrac{\mathrm{S_{file}}}{T_\mathrm{depature}-T_\mathrm{arrival}}&, \text{Transmitted File}\\
	\dfrac{S_{\mathrm{served}}}{T_\mathrm{end}-T_\mathrm{arrival}}& ,\text{Unfinished File}\\
	0& ,\text{Dropped File}
	\end{cases}
	\end{equation}
	where $\mathrm{S_{file}}$ is the file size, $S_{\mathrm{served}}$ is the transmitted file size by the end of simulation,   $T_\mathrm{arrival}$ is the arrival time of the file, and $T_\mathrm{depature}$ represents the transmitted time of the file. It is noted that the file throughput is zero when the file is dropped due to time violation.
	
	\subsection{Hidden Node and Exposed Node Problem}
		As the devices contend for the unlicensed spectrum without coordination, in both NR-U and WiFi networks, each device is required to perform Clear Channel Assessment (CCA) to sense the ongoing transmissions over the channel before its own transmission, where the sensed energy is compared with the fixed Energy Detection (ED) threshold to determine whether the channel is idle or busy. In other words, each device can only transmit when the channel is detected to be idle. However, the CCA procedure based on the fixed ED threshold cannot fully mitigate the throughput degradation due to the lack of coordination function among devices, which leads to the hidden node (HN) problem and exposed node (EN) problem \cite{Sun2020}.

\begin{figure}[!h]
	\centerline{\includegraphics[scale=0.5]{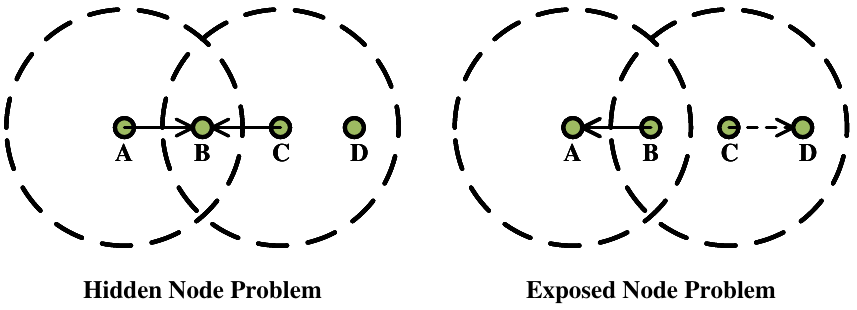}}
	\caption{Hidden node problem and exposed node problem.}
	\label{hidden_exposed_node}
\end{figure}

As shown in the Fig.~\ref{hidden_exposed_node}, the HN problem happens when transmitter A and transmitter C cannot sense the transmission of each other (i.e., the sensed energy is lower than the ED threshold) due to the pathloss and fading. Hence, two transmitters will send the packets simultaneously, which causes strong co-channel interference and transmission failure at receiver B. The EN problem happens when the transmitter C is prevented from sending packet to receiver D because of co-channel interference from the neighboring transmitter B (i.e., the sensed energy is higher than the ED threshold). However, receiver D could still receive the transmission from C with low interference because receiver D is far away from transmitter B, which leads to low spectral efficiency. Moreover, NR-U and WiFi networks adopt asymmetry ED thresholds (e.g., -72dBm in NR-U network and -62dBm in WiFi network), which makes the HN and EN problems even worse \cite{Mehrnoush2018}. In a word, the high ED threshold setting enables multiple devices to transmit simultaneously, but may lead to stronger interference. On the contrary, the low ED threshold setting mitigates the probability of collision, but leads to low spectrum efficiency \cite{Yang2018}.

	\subsection{Uplink System Throughput Optimization}
	As discussed above, the WiFi STAs perform CSMA/CA individually to access the channel without scheduling, and all the STAs that accessed the channel would transmit in a grant-free manner. The gNB is required to perform the Cat4 LBT to access the channel, and transmits the scheduling information via the PDCCH. Then, the scheduled UE is required to perform Cat2 or Cat4 LBT to transmit the user data based on its scheduled resource. In this work, we tackle the problem of optimizing the coexistence of NR-U and WiFi configuration defined by action parameters $A^{t} = \left\lbrace \lambda^{\mathrm{ED},t}_{\mathrm{w}} ,\lambda^{\mathrm{ED},t}_{\mathrm{n}} \right\rbrace $ for  WiFi and NR-U networks in each transmission time interval (TTI) $t$, where $\lambda^{\mathrm{ED},t}_{\mathrm{w}}$ and $\lambda^{\mathrm{ED},t}_{\mathrm{n}}$ represent the ED threshold for NR-U and WiFi networks, respectively. 
	
	In order to select the action at the beginning of every TTI $t$, the central server accesses all prior historical observations $U^{t^{'}}$ in TTIs $ t^{'} =1,...t-1$ consisting of the following variables: the number of successfully transmitted WiFi packets $N_{\mathrm{w}}^{t^{'}}$, and the number of successfully transmitted NR-U packets $N_{\mathrm{n}}^{t^{'}}$. We denote $O^{t}=\left\lbrace A^{t-1}, U^{t-1},A^{t-2}, U^{t-2},...,A^{1}, U^{1}\right\rbrace $ as the observed history of all such measurements and past actions.
	
	The WiFi and NR-U networks aim at maximizing the long-term average system throughput with respect to the stochastic policy $\pi$ that maps the current observation history $O^{t}$ to the probabilities of selecting each possible configuration $A^{t}$. This problem can be formulated as
	\begin{equation} (\text {P1})\!:\quad \mathop {\textbf {max}}\limits _{ \{\pi (A^{^{t}}|O^{^{t}})\}}~\sum _{k=t}^{\infty } \gamma ^{(k-t)} {\mathbb E}_{\pi } [\left( N_{\mathrm{w}}^{k} + N_{\mathrm{n}}^{k}\right)\times\mathrm{S_\mathrm{file}} ],
	\end{equation}
	where $\gamma \in [0,1)$ is the discount rate for the performance in future TTIs, and $\mathrm{S_\mathrm{file}}$ is the file size. Since the dynamics of the system is Markovian over the TTIs, this is a partially observable Markov decision process (POMDP) problem that is generally intractable. Approximate centralized and federated solutions will be discussed in Section \ref{centralized} and Section \ref{dsitributed}, respectively.
	
	\section{Uplink System Throughput Optimization based on Deep Reinforcement Learning}
	\label{DRL_chapter}
	DRL is one of the most promising solutions to optimally solve complex POMDP problems, due to the reliance on the deep neural networks as one of the most powerful non-linear approximation functions \cite{Barto2018}. The related DRL algorithms have been widely used in the dynamic optimization for wireless communication systems, e.g., \cite{jiang2019,jiang2021,Liu2022}. However, they have been rarely studied in handling the ED thresholds configuration under the coexistence of heterogeneous NR-U and WiFi networks. Therefore, we design the centralized DRL and federated DRL in this section to evaluate the capability of DRL algorithm, respectively.
	
	\subsection{Centralized Deep Reinforcement Learning}
	\label{centralized}
	
	\begin{figure*}[!h]
		\centerline{\includegraphics[scale=0.5]{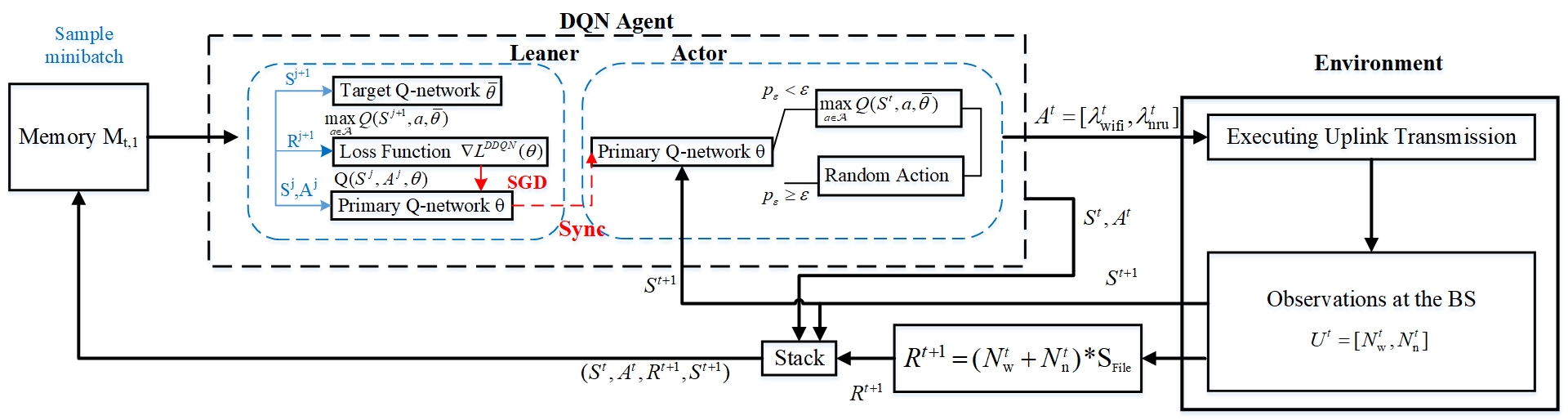}}
		\caption{The DQN agent and environment interaction in the POMDP.}
		\label{DQN_eMBB}
	\end{figure*}
	
	In the central server, an agent is deployed to optimize the system throughput of the NR-U and WiFi networks, the agent is required to explore the environment in order to choose appropriate actions progressively leading to the optimization goal. We define $s \in \mathcal{S}$, $a \in \mathcal{A}$, and $r \in \mathcal{R}$ as any state action, and reward from the their corresponding sets, respectively. At the beginning of the $t$th TTI $\left(t\in \left\lbrace0,1,2,...\right\rbrace\right)$, the agent first observes the current state $S^{t}$ corresponding to a set of previous observations $\left(O^{t}=\left\lbrace U^{t-1}, U^{t-2}, ..., U^{1}\right\rbrace\right)$ in order to select an specific action $A^{t}\in \mathcal{A}\left(S^{t}\right)$. The action $A^{t}$ is designed to be the WiFi and NR-U ED thresholds $\lambda^{\mathrm{ED},t}_{\mathrm{w}} ,\lambda^{\mathrm{ED},t}_{\mathrm{n}} $, respectively.
	
	We consider a basic state function under the coexistence of NR-U and WiFi networks, where $S^{t}$ is a set of indices mapping to the current observed information $U^{t-1}=\left[N_{\mathrm{w}}^{t-1}, N_{\mathrm{n}}^{t-1}\right]$. With the knowledge of the state $S^{t}$, the agent chooses an action $A^{t}$ from the set $\mathcal{A}$, which is a set of indices mapped to the set of available ED thresholds $\bm{\mathcal{F}}_\mathrm{ED}$. Once an action $A^{t}$ is performed, the agent will receive a scalar reward $R^{t+1}$, and observe a new state $S^{t+1}$. The reward $R^{t+1}$ indicates to what extent the executed action $A^{t}$ can achieve the optimization goal, which is determined by the new observed state $S^{t+1}$. As the optimization goal is to maximize the system throughput of NR-U and WiFi networks, we define the reward $R^{t+1}$ as the size of transmitted data packets in each TTI
		\begin{equation}
			R^{t+1}=(N_{\mathrm{w}}^{t}+N_{\mathrm{n}}^{t})*\mathrm{S_{file}}.
			\label{eq:centralized_reward}
		\end{equation}
		where $\mathrm{S_{file}}$ is the size of each packet, $N_{\mathrm{w}}^{t-1}$ is number of successfully transmitted packets in the WiFi network, and $N_{\mathrm{n}}^{t-1}$ is number of successfully transmitted packets in the NR-U network.
	
	The basic Q-learning is a value-based RL approach, where the policy of states to actions mapping $\pi\left(s\right)=a$ is learned using a state-action value function $Q\left(s,a\right)$ to determine an action for the state $s$. In the Q-learning, a lookup table is utilized to represent the state-action value function $Q\left(s,a\right)$, which consists of value scalars for all the state and action spaces. To obtain an action $A^{t}$, the highest value scalar is selected from the numerical value vector $Q\left(S^{t},a\right)$, which maps all possible actions under $S^{t}$ to the Q-value table $Q\left(s,a\right)$.
	
	Accordingly, the objective is to find an optimal Q-value table $Q^{*}\left(s,a\right)$ with optimal policy $\pi^{*}$ that can select actions to dynamically optimize the system throughput of NR-U and WiFi networks. To do so, an initial Q-value table $Q\left(s,a\right)$ is trained in the environment, where $Q\left(s,a\right)$ is immediately updated using the current observed reward $R^{t+1}$ after each action as
	\begin{equation}
	\begin{aligned}
	&Q\left(S^{t},A^{t}\right)=Q\left(S^{t},A^{t}\right)+\\&\eta\left[R^{t+1}+\gamma \mathop{max}\limits_{a\in \mathcal{A}}Q\left(S^{t+1},a\right)-Q\left(S^t,A^t\right)\right],
	\end{aligned}
	\label{eq:Q_learning}
	\end{equation}
	where $\eta$ is a constant step-size learning rate that affects how fast the algorithm adopts to a new environment, $\gamma \in [0,1)$ is the discount rate that determines how current rewards affects the value function updating $\mathop{max}\limits_{a\in \mathcal{A}}Q\left(S^{t+1},a\right)$ approximates the value in optimal Q-value table $Q^*(s,a)$ via the up-to-date Q-value table $Q(s,a)$ and the obtained new state $S^{t+1}$.

	Specifically, the learning rate $\eta$ is suggested to be set to a small number (e.g., $\eta=0.01$) to guarantee the stable convergence of Q-value table under the coexistence of NR-U and WiFi networks. This is due to that a single reward in a specific TTI can be severely biased, because state function is composed of multiple unobserved information with unpredictable distributions.

	% Generally, selecting an efficient approximation approach to represent the value function for different learning scenarios is a usual problem within the RL. A variety of function approximation approaches can be conducted, such as LA, DNNs, tree search, and which approach to be selected can critically influence the successful learning. The function approximation should fit the complexity of the desired value function, and be efficient to obtain good solutions. Unfortunately, most function approximation approaches require specific design for different learning problems, and there is no basis function, which is both reliable and efficient to satisfy all learning problems.

	%The reasons we conduct DQN are that: 1) the DNN function approximation is able to deal with several kinds of partially observable problems; 2) DQN has the potential to accurately approximate the desired value function while addressing a problem with very large state spaces, which can be favored for the learning in the multiple CE group scenarios; 3) DQN is with high scalability, where the scale of its value function can be easily fit to a more complicated problem; 4) a variety of libraries have been established to facilitate building DNN architectures and accelerate experiments, such as TensorFlow, Pytorch, Theano, Keras, and etc..
	
	However, Q-learning needs each element to be updated to converge, searching for an optimal policy can be difficult with limited time and computational resources. To solve this problem, we use a value function approximator instead of Q-value table to find a sub-optimal approximated policy. We conduct the DNN for Q-learning as a more effective but complicated function approximator, which is also known as DQN. As shown in the Fig.~\ref{DQN_eMBB}, the DQN agent parameterizes the action-state value function $Q(s, a)$ by using a function $Q(s, a; \theta)$, where $\theta$ represents the weights matrix of a DNN with multiple layers. In the learner, the weights $\theta$ is updated through Stochastic gradient descent (SGD) based on the minibatch sample. In the actor, the action is determined based on the $\epsilon$-greedy approach. We consider the conventional DNN, where neurons between two adjacent layers are fully pairwise connected, namely fully-connected layers. The input of the DNN is given by the variables in state $S^{t}$; the intermediate hidden layers are Rectifier Linear Units (ReLUs) by using the function $f(x) = max (0, x)$; while the output layer is composed of linear units, which are in one-to-one correspondence with all available actions in $\mathcal{A}$.
	
	We consider $\epsilon$-greedy approach to balance exploitation and exploration in the Actor of the Agent, where $\epsilon$ is a positive real number and 
		$\epsilon\leq 1$. In each TTI $t$, the agent randomly generates a probability $p^{t}_{\epsilon}$ to compare with $\epsilon$. Then, with the probability $\epsilon$, the algorithm randomly chooses an action from the remaining feasible actions to improve the estimate of the non-greedy action’s value. With the probability $1-\epsilon$, the exploitation is obtained by performing forward propagation of Q-function $Q(s, a; \theta)$ with respect to the observed state $S^{t}$, which can be calculated as follows:
		\begin{equation}
			a=\begin{cases}
				\mathop {\text {max}}\limits _{a\in A}Q(S^{i+1}, a; \bar {\boldsymbol {\theta }}^{t})&, \text{with probability 1 - $\epsilon$}\\
				\mathrm{Random~action}& ,\text{with probability $\epsilon$}.
			\end{cases}
	\end{equation}
	
	The weights matrix $\theta$ is updated online along each training episode by using double deep Q-learning (DDQN), which to some extent reduce the substantial overestimations of value function. Accordingly, learning takes place over multiple training episodes, with each episode of duration $N_{\mathrm{TTI}}$ TTI periods. In each TTI, the parameter $\theta$ of the Q-function approximator $Q(s, a; \theta)$ is updated using RMSProp optimizer as
	\begin{equation} 
	{\boldsymbol \theta }^{t+1} = {\boldsymbol \theta }^{t} - \lambda _{\text {RMS}} \nabla L({\boldsymbol \theta }^{t}),
	\label{eq:optimizer}
	\end{equation}
	where $\lambda _{\text {RMS}}$ is RMSProp learning rate, $\nabla L({\boldsymbol \theta })$ is the gradient of the loss function $L({\boldsymbol \theta })$ used to train the Q-function approximator. This is given as
	\begin{equation} 
	\begin{aligned}
	\nabla L(\boldsymbol{\theta} ^{t})=&{\mathbb E}_{S^{i},A^{i},R^{i+1},S^{i+1}} \big [\!\big (\!R^{i+1}\!+\! \gamma \mathop {\text {max}}\limits _{a}Q(S^{i+1}, a; \bar {\boldsymbol {\theta }}^{t}) \\&\qquad \,\!-\,Q(S^{i}, A^{i}; \boldsymbol {\theta }^{t}) \big) \nabla _{\boldsymbol {\theta }} Q(S^{i}, A^{i}; \boldsymbol {\theta }^{t})\big],
	\label{eq:DQN_update}
	\end{aligned}
	\end{equation}
	where the expectation is taken with respect to a so-called minibatch, which are randomly selected previous samples $(S^{i},A^{i},S^{i+1},R^{i+1})$ for some $i \in {t-M_r,...,t}$, with $M_r$ being the replay memory. When $t-M_r$ is negative, this is interpreted as including samples from the previous episode. The use of minibatch, instead of a single sample, to update the value function $Q(s, a; \theta)$ improves the convergent reliability of value function. Furthermore, following DDQN, in \eqref{eq:DQN_update}, $\bar {\boldsymbol {\theta }}^{t}$ is a so-called target Q-network that is used to estimate the future value of the Q-function in the update rule. This parameter is periodically copied from the current value $\boldsymbol {\theta }^{t}$ and kept fixed for a number of episodes. In the following, the implementation of Centralized-DQN based ED thresholds configuration is shown in \textbf{Algorithm \ref{Centralized_DQN}} .
	
		It is noted that Double DQN algorithm outperforms the Q-table and DQN algorithm as demonstrated in \cite{VanHasselt2016}. Therefore, we focus on the Double DQN algorithm in the paper. 
	\begin{algorithm}
		\caption{Centralized-DQN Based Energy Detection Thresholds Configuration}
		\label{Centralized_DQN}
		\LinesNumbered 
		\KwIn{The set of available energy thresholds $\bm{\mathcal{F}}_\mathrm{ED}$ under the coexistence of WiFi and NR-U networks.}
		Algorithm hyperparameters: learning rate $\lambda_{\text {RMS}} \in (0,1]$, discount rate $\gamma \in (0,1]$, $\epsilon$-greedy rate $\epsilon \in (0,1]$, target network update frequency $K$\;
		
		Initialization of replay memory $M$ to capacity $C$, the primary Q-network $\boldsymbol{\theta}$, and the target Q-network $\boldsymbol{\overline{\theta}}$\;
		
		\For{episode $= 1,..,I$}{
			Initialization of $S^{1}$ by executing a random action $A^{0}$\;
			\For{$t= 1,..,T$}{		
				Update the traffic\;
				\eIf{$p_{\epsilon}<\epsilon$}{
					select a random action $A^{t}$ from $\mathcal{A}$\;
				}{
					select $A^{t}=\mathrm{argmax}Q(S^{t},a,\boldsymbol{\theta})$\;
				}
				The central server broadcasts the ED thresholds to all the devices\;
				The central server observes $S^{t+1}$, and calculate the related $R^{t+1}$ using Eq.~\ref{eq:centralized_reward}\;
				Store transition $(S^{t}, A^{t}, R^{t+1}, S^{t+1})$ in replay memory $M$\;
				Sample random minibatch of transitions $(S^{t}, A^{t}, R^{t+1}, S^{t+1})$ from replay memory $M$\;
				Perform a gradient descent for $Q(s,a,\boldsymbol{\theta})$ using \ref{eq:DQN_update}\;
				Every $K$ setps update target Q-network $\overline{\boldsymbol{\theta}}=\boldsymbol{\theta}$.
			}
		}
	\end{algorithm}

	\subsection{Federated Deep Reinforcement Learning}
	\label{dsitributed}
	In the centralized DRL, the agent needs to access the number of successfully transmitted packets both from NR-U and WiFi networks during the training and testing phases. However, it is usually impractical for heterogeneous NR-U and WiFi networks to share this information due to data privacy regulation and network architecture \cite{Sylla2022}. Therefore, we propose  federated DRL in the optimization of ED thresholds to solve this problem.
	
	%FL is a machine learning setting where many clients (e.g. mobile devices or whole organizations) collaboratively train a model under the orchestration of a central server (e.g. service provider), while keeping the training data decentralized. It embodies the principles of focused collection and data minimization, and can mitigate many of the systemic privacy risks and costs resulting from traditional, centralized machine learning. This area has received significant interest recently, both from research and applied perspectives.

	\begin{figure}[!h]
		\centerline{\includegraphics[scale=0.5]{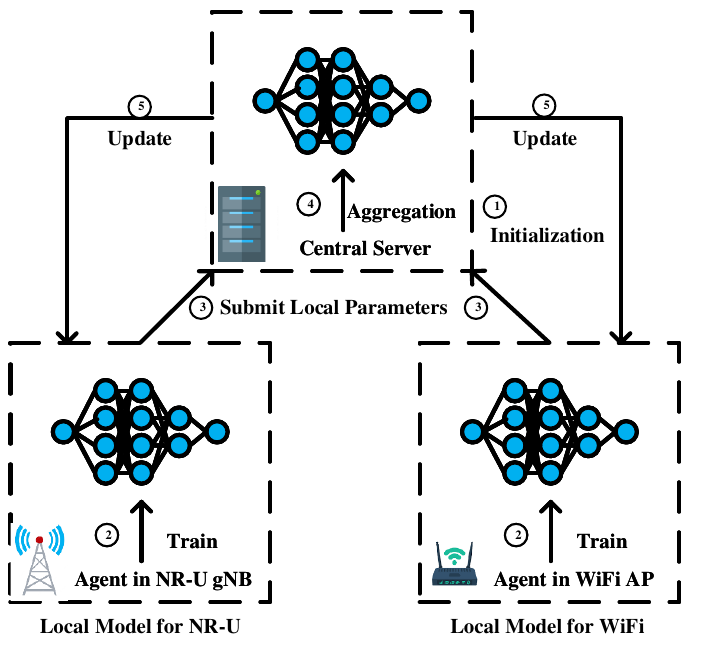}}
		\caption{Structure of federated deep reinforcement learning.}
		\label{Federated_eMBB}
	\end{figure}

	As shown in the Fig.~\ref{Federated_eMBB}, in the federated DRL, two independent agents are deployed in the NR-U and WiFi networks, respectively. Each agent only observes the number of successfully transmitted packets of its own system. During training, each agent reports the DQN parameters to the central server, and the central server calculates the averaged global parameters. Then each agent updates the parameters with the received global parameters. The basic workflow of federated DRL can be summarized in the following five steps:
	\begin{itemize}
		\item Step 1: : Initialization is performed for both federated DRL agents
		models.
		\item Step 2: : Each federated DRL agent trains a local model based on
		their respective dataset following \textbf{Algorithm \ref{Centralized_DQN}}. 
		\item Step 3: : Each federated DRL agent reports its model parameters
		$({\pmb {\theta }}_{n}, {\pmb {\theta }}_{w})$ to the central server.
		\item Step 4: : The central server combines the model parameters
		via specific aggregation.
		\item Step 5: : The combined global parameter $\pmb {\theta}$ is
		sent back to the corresponding agents and both agents
		update their local models based on the received global
		parameters.
\end{itemize}
	
 During training, the parameter 
	$\widetilde {\pmb {\theta }}_l,l=\{n,w\}$ can be updated as
	\begin{equation} 
	\widetilde {\pmb {\theta }}_{l}^{t+1} = \widetilde {\pmb {\theta }}_{l}^{t} - \lambda _{\text {RMS}}\nabla L_{l}(\widetilde {\pmb {\theta }}_{l}^{t}),
	\end{equation}
	where $\lambda_{\text {RMS}} \in$(0,1] is the learning rate, $\nabla L_{l}(\widetilde {\pmb {\theta }}_{l}^{t})$ is the loss function gradient of the NR-U network or the WiFi network, which can be denoted as
	\begin{equation} 
	\begin{aligned}
	\nabla L_l(\boldsymbol{\theta} ^{t}_l)=&{\mathbb E}_{S^{i}_l,A^{i}_l,R^{i+1}_l,S^{i+1}_l} \big [\!\big (\!R^{i+1}_l\!+\! \gamma \mathop {\text {max}}\limits _{a}Q(S^{i+1}_l, a; \bar{\widetilde {\boldsymbol {\theta }}}^{t}_l ) \\&\qquad \,\!-\,Q(S^{i}_l, A^{i}_l; \widetilde {\boldsymbol {\theta }}^{t}_l) \big) \nabla _{\boldsymbol {\theta }_l} Q(S^{i}_l, A^{i}_l; \widetilde{\boldsymbol {\theta }}^{t}_l)\big],
	\end{aligned}
	\end{equation}
	where $(S_l^i,A_l^i,S_l^{i+1},R_l^{i+1})$ are randomly selected previous samples for $i \in \{t-M_\mathrm{r},...,t\}$. $\bar {\widetilde {\pmb {\theta }}}_{l}^{t}$ is the target Q-network which is used to estimate the future value of the state-action value function in the update rule. In addition, the updated parameter $\widetilde {\pmb {\theta }}_l$ will be transmitted to the central server and the model parameter $\boldsymbol{\theta}$ can be updated as
	\begin{equation} \pmb {\theta } = \frac{1}{2}\left( \widetilde {\pmb {\theta }}_{n}+\widetilde {\pmb {\theta }}_{w}\right),
	\end{equation}
	where $\widetilde {\pmb {\theta }}_{n}$ and $\widetilde {\pmb {\theta }}_{w}$ represent the model parameters in the NR-U and WiFi networks, respectively.
	
	It is noted that Gate Recurrent Unit (GRU) is adopted in the proposed learning framework to approximate the value function. The complexity of GRU is calculated as $O(n_i n_l n_{h}^2)$, where $n_i$ is the input size, $n_l$ is the number of layers, and $n_h$ is the hidden size. Without loss of generality, we assume the number of training episodes for convergence to be $I$, and length of each episode to be $T$. Therefore, the complexity of centralized-DQN and federated-DQN is $O(n_i n_l n_{h}^2 I T) $ and $ O(X n_i n_l n_{h}^2 I T) $, respectively, where $X$ is the number of agents in the federated-DQN.

	\section{Simulation}
	In this section, we evaluate the performance of the proposed centralized DQN and federated DQN approaches in Sec.~\ref{DRL_chapter} via numerical experiments. We adopt the standard network parameters listed in Table.~\ref{Simulation_parameter} following \cite{channel_model_2019,LAA2015,NRU2018,TrafficFTP32017, Li2017}. In the following, we first present our simulation results of system throughput optimization in Section \ref{uplink_thoughput_optimization_section}, then redesign the reward function to guarantee the cell throughput of WiFi network with more dynamic Beta traffic in Section \ref{uplink_thoughput_optimization_section_with_fairness_beta}.
	\begin{table}[h!]
		\caption{Simulation Parameters}
		\begin{center}
			\begin{tabular}{|c|c|}
				\hline 
				\rowcolor{gray}\textbf{Parameters} & \textbf{Value} \\
				\hline
				Height of gNB and AP & $3$ m  \\ 
				\hline
				Height of UE and STA& $1$ m \\
				\hline
				File Size& 0.5 Mbytes \\
				\hline
				SIFS& 16us \\
				\hline
				Defer& 79us \\
				\hline
				Maximum Contention Window& 1024 \\
				\hline
				Packet Arrival Rate& 2 \\
				\hline
				Transmit Power of UE and STA& 18dBm \\
				\hline
				Transmit Power of gNB and AP& 23dBm \\
				\hline
				Noise Power& -104dBm \\
				\hline
				SCS& 60KHz\\
				\hline
				Mini-slot&36us\\
				\hline
				Time Limitation& 8s \\
				\hline	
				WiFi SINR Threshold& 9dB \\
				\hline
				NR-U SINR Threshold& 5.5dB \\
				\hline	
				WiFi Rate& 21.7Mbps \\
				\hline
				NR-U Rate & 25.2Mbps \\
				\hline	
				WiFi COT& 2.528ms \\
				\hline
				NR-U COT & 6ms \\
				\hline
				WiFi Preamble Detection Threshold& -82dBm \\
				\hline
				WiFi Energy Detection Threshold & -62dBm \\
				\hline
				NR-U Energy Detection Threshold & -72dBm \\
				\hline
				Number of UEs associated with each gNB & 5 \\
				\hline
				Number of STAs associated with each AP & 5 \\
				\hline
				Simulation Length& 250s \\
				\hline
				
			\end{tabular}
			\label{Simulation_parameter}
		\end{center}
	\end{table}

	\subsection{Uplink System Throughput Optimization}
	\label{uplink_thoughput_optimization_section}
	
	\begin{figure}[!h]
		\centerline{\includegraphics[scale=0.4]{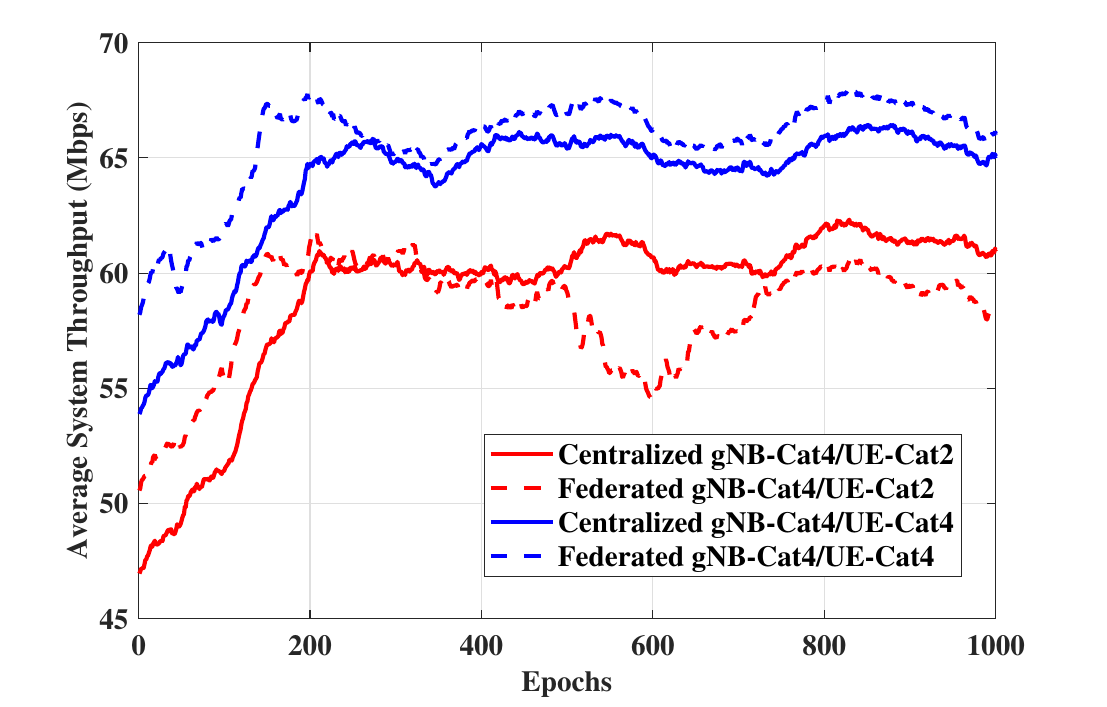}}
		\caption{Average uplink system throughput during training phase of proposed DRL algorithms.}
		\label{Convergence}
	\end{figure}

	\begin{figure*}[!h]
	\centering
	\subfigure[Centralized DRL with gNB-Cat4/UE-Cat2 LBT]
	{\includegraphics[scale=0.4]{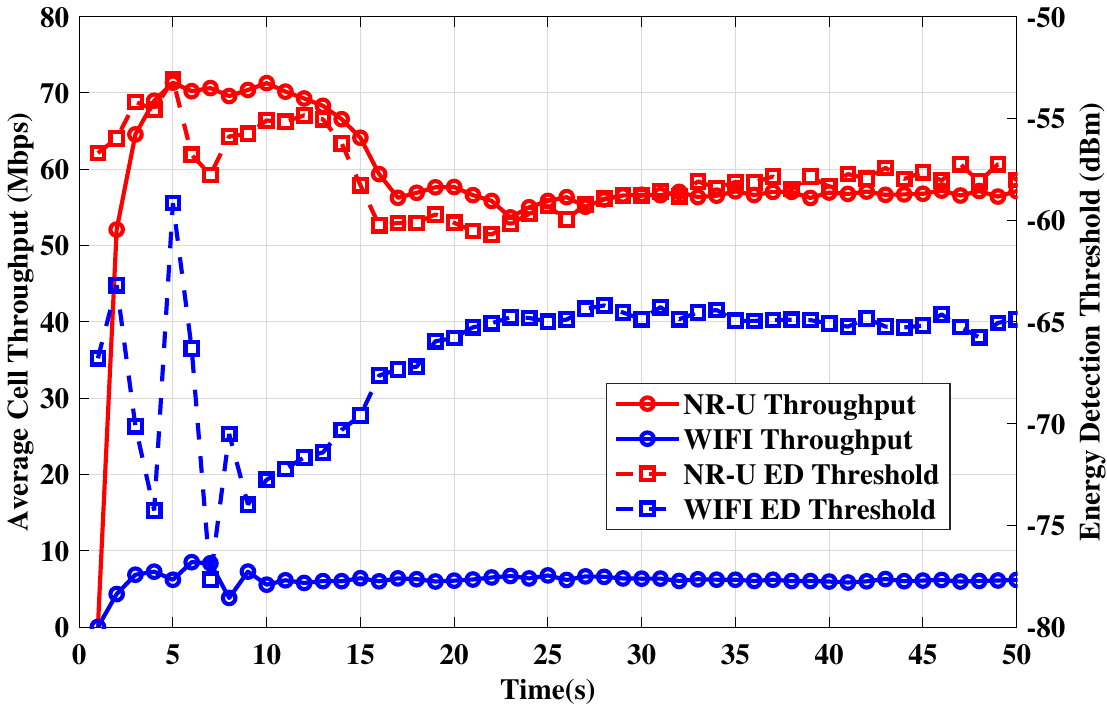}} 
	%\\
	%\centering
	\subfigure[Federated DRL with gNB-Cat4/UE-Cat2 LBT]
	{\includegraphics[scale=0.4]{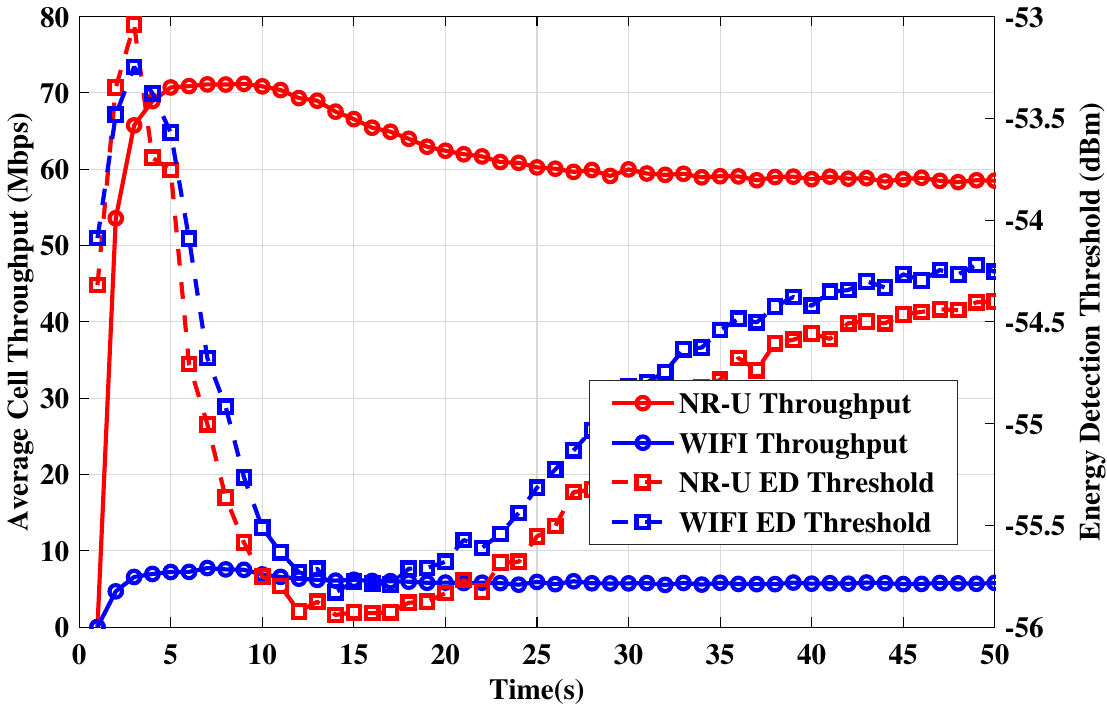}}\\
	\subfigure[Centralized DRL with gNB-Cat4/UE-Cat4 LBT]
	{\includegraphics[scale=0.4]{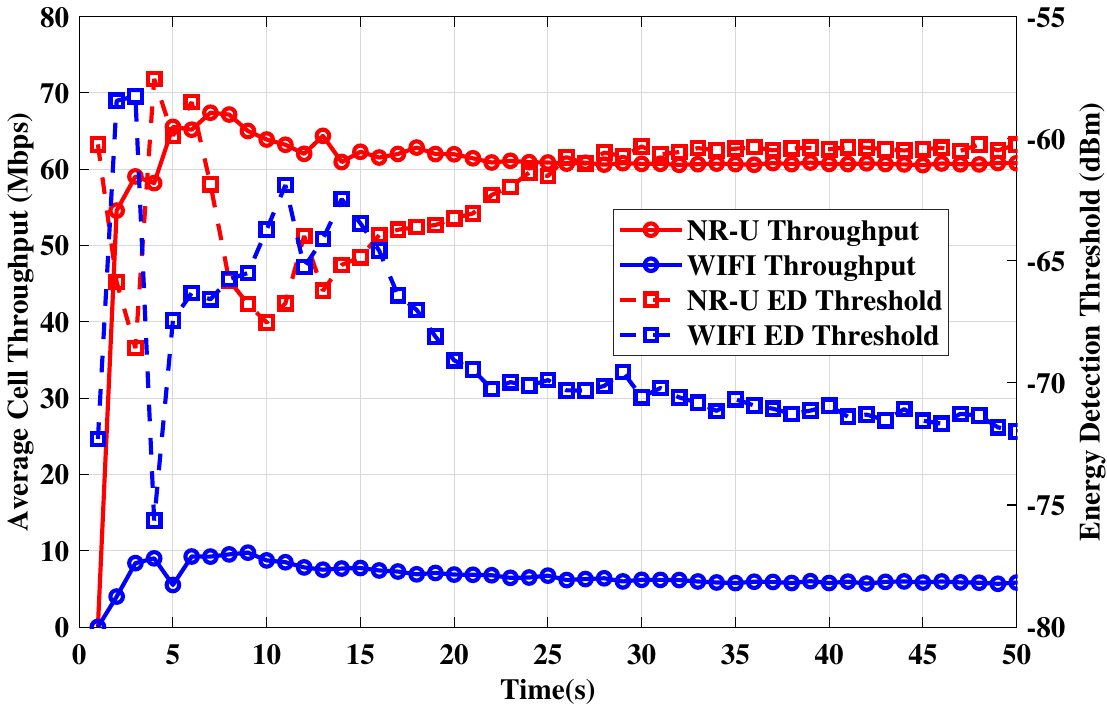}} 
	%\\
	%\centering
	\subfigure[Federated DRL with gNB-Cat4/UE-Cat4 LBT]
	{\includegraphics[scale=0.4]{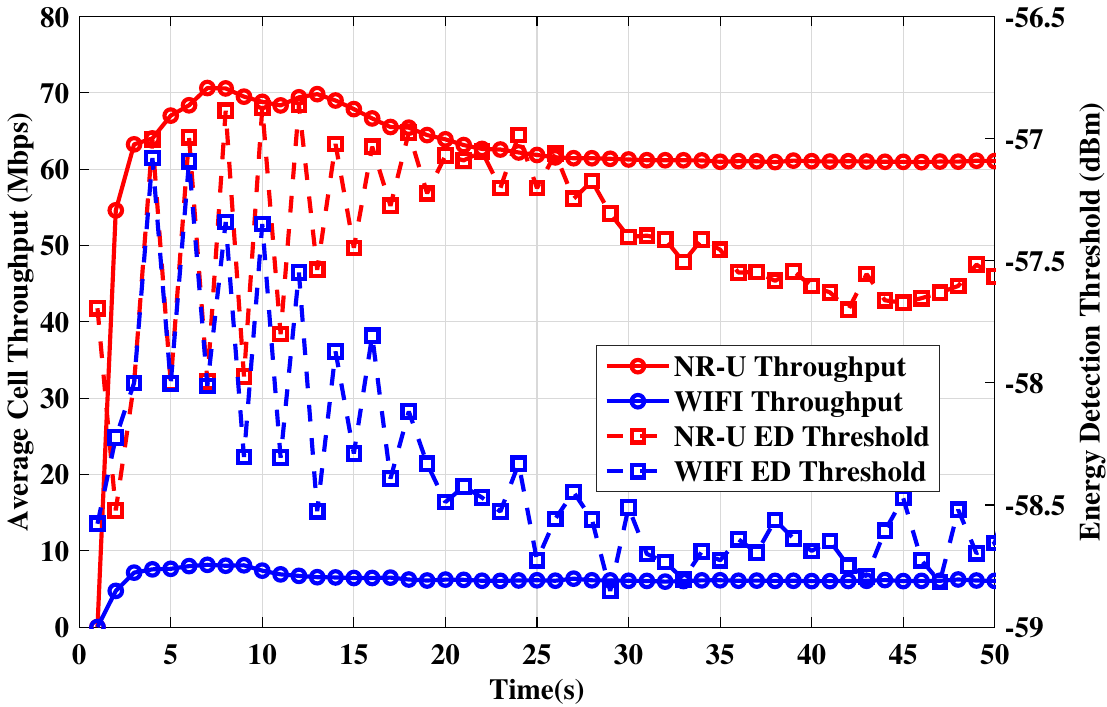}}\\
	\caption{Cell throughput and actions of WiFi and NR-U networks during testing phase.}
	\label{Testing}
\end{figure*}

	Fig.~\ref{Convergence} plots the average uplink system throughput during the training phase of proposed centralized and federated DRL algorithms. We can see that the average uplink system throughput of proposed centralized and federated DRL algorithms with gNB-Cat4/UE-Cat4 reaches around 65Mbps. However, the average uplink system throughput with gNB-Cat4/UE-Cat2 stabilizes only around 60Mbps. This is because the gNB has to reschedule the UE when the channel is busy during Cat2 LBT, which leads to unnecessary controlling overheads. We can also observe that the federated DRL algorithms converge faster than the centralized DRL algorithms due to the much smaller action space in the setting. More importantly, centralized-DQN may easily converge to the local optimal solution due to large action space. Therefore, we can observe that the federated-DQN achieves higher performance than the centralized-DQN algorithm. However, the training phase of federated DRL algorithms is more fluctuated  due to the model parameters averaging between heterogeneous networks.
	
	%We can also observe that the average system throughput of federated DRL with gNB-Cat4/UE-Cat2 converges around 1200 episodes, which is much slower compared to 250 episodes in centralized DRL. Moreover, the average system throughput of federated DRL with gNB-Cat4/UE-Cat4 converges around 800 episodes, which is also slower compared to 250 episodes in centralized DRL. This is because federated learning needs more time to learn from each agent with only local observation, whereas centralized learning can learn from global observations. 

	As shown in Fig.~\ref{Testing}, we test the performance of converged trained model, where the average cell throughput and ED thresholds of NR-U and WiFi networks are characterized, respectively. We can see that the networks reach the stable state around 40 seconds. This is because the traffic follows Poisson arrival with a fixed average arrival rate, and the packet that violates the time constraint is dropped. 
	
	Fig.~\ref{Testing} (a)(b) plot the average cell throughput and ED thresholds with gNB-Cat4/UE-Cat2 LBT scheme. In Fig.~\ref{Testing}(a), we can observe that the cell throughput of NR-U network is around 58 Mbps, which is much higher than 6 Mbps of WiFi network. This can be explained from two aspects: 1) NR-U network has advantages over WiFi network including higher data rate, larger COT length, centralized scheduling, etc; 2) Due to the advantages of NR-U network, it can contribute more significantly to the improvement of the system throughput. Therefore, we can see that NR-U network adopts the ED threshold around -57 dBm, which is higher than the -66 dBm of WiFi network. The higher ED threshold further enables the NR-U network to get more transmission opportunities. In Fig.~\ref{Testing}(b), we can see that although the NR-U and WiFi networks adopt similar ED thresholds around -54 dBm, the uplink throughput of NR-U and WiFi networks are still around 58 Mbps and 6 Mbps, respectively. The reason is that there is the other CS threshold $\lambda^{\mathrm{CS}}_{\mathrm{w}}=-82 \mathrm{dBm}$, which limits the transmission of the WiFi network. 
	
	Fig.~\ref{Testing} (c)(d) plot the average cell throughput and ED thresholds with gNB-Cat4/UE-Cat4 LBT scheme. Compared with gNB-Cat4/UE-Cat2 LBT scheme, we can see that the NR-U network achieves higher cell throughput around 61 Mbps, while the cell throughput of WiFi network is still around 6 Mbps. This is because the gNB is required to perform the Cat4 LBT to schedule the UE again when the UE detects the channel busy during Cat2 LBT. However, if the UE performs the Cat4 LBT, the UE can adopt the Backoff procedure to wait for the transmission opportunity, which leads to less controlling overheads. We can also observe that the ED thresholds of both centralized and federated DRL algorithms are lower than that of gNB-Cat4/UE-Cat2 LBT scheme. The reason is that the gNB-Cat4/UE-Cat-4 scheme can deal with the traffic more effectively, hence, the interference level is relatively lower.

	\begin{figure}[!h]
		\centerline{\includegraphics[scale=0.45]{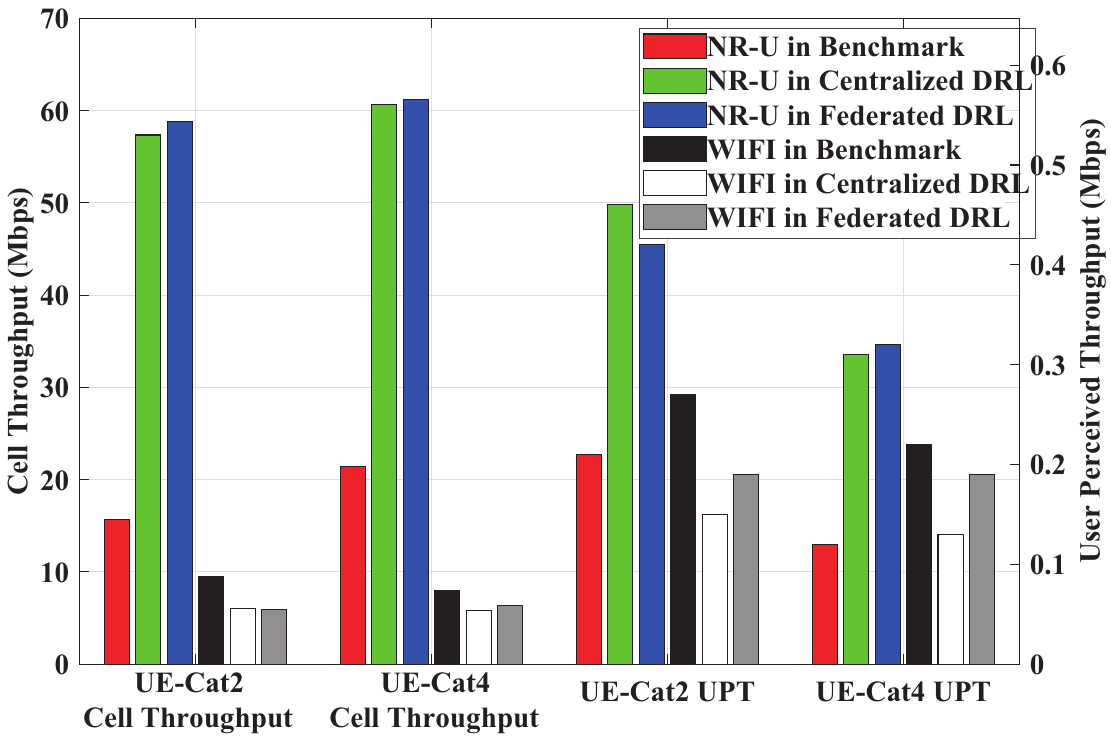}}
		\caption{Average cell throughput and user perceived throughput during testing phase.}
		\label{UPT}
	\end{figure}
	
	Fig.~\ref{UPT} plots the average cell throughput and UPT of the proposed DRL algorithms and benchmark scheme with fixed ED thresholds. In the gNB-Cat4/UE-Cat2 scheme, we can see that the cell throughput of NR-U network increases from 16 Mbps to 58 Mbps with over 250\% performance gain, and the cell throughput of WiFi network decreases from 9.5 Mbps to 6 Mbps with performance degradation around 35\%. The system throughput increases from 25.5 Mbps to 64 Mbps, where the performance gain is over 150\%. The performance gain of the NR-U network comes from the much higher ED threshold compared to the WiFi network as shown in Fig.~\ref{Testing}, which indicates that the agent chooses to sacrifice the performance of WiFi network to get higher system throughput. 
	
	We can observe that the UPT of NR-U network in the benchmark is lower than that of WiFi network due to the scheduling overheads in the NR-U network. In the gNB-Cat4/UE-Cat2 scheme, the UPT of NR-U network increases from 0.21 Mbps to around 0.45 Mbps with performance gain over 100\%, and UPT of WiFi network decreases from 0.27 Mbps to around 0.17 Mbps with performance degradation around 35\%. This is because the NR-U network can occupy the channel to transmit the packet more timely with a higher ED threshold. It is noted that the UPT of NR-U network in federated DRL is lower than that of centralized learning. However, the UPT of WiFi network in federated DRL is higher than that of centralized learning. The reason is that both NR-U and WiFi networks adopt higher ED thresholds in federated DRL. The interference level experienced by the NR-U network reaches its decoding capability limitation due to the high ED threshold. On the contrary, the interference level experienced by the WiFi network is far away from its decoding capability limitation due to the CS threshold $\lambda^{\mathrm{CS}}_{\mathrm{w}}=-82 \mathrm{dBm}$.
	
	In the gNB-Cat4/UE-Cat4 scheme, we can see that the cell throughput of NR-U network increases from 21 Mbps to 61 Mbps with over 180\% performance gain, and the cell throughput of WiFi network decreases from 8 Mbps to 6 Mbps with performance degradation around 20\%. The system throughput increases from 29 Mbps to 67 Mbps, where the performance gain is over 120\%. The UPT of NR-U network increases from 0.12 Mbps to around 0.31 Mbps with performance gain over 150\%, and UPT of WiFi network decreases from 0.27 Mbps to around 0.16 Mbps with performance degradation around 20\%. As we discussed above, since the interference level is lower in the gNB-Cat4/UE-Cat4 scheme, we can see that the UPT of both NR-U and WiFi networks in federated DRL is higher than that of centralized DRL. This is because the interference with higher ED thresholds is still within the decoding capability of NR-U and WiFi devices.
	
	\subsection{Uplink System Throughput Optimization with Fairness}
	\label{uplink_thoughput_optimization_section_with_fairness_beta}
	In Section \ref{uplink_thoughput_optimization_section}, the simulation results have shown that although the system throughput has been improved significantly, the agent chooses to sacrifice the cell throughput of the WiFi network to get higher cell throughput of NR-U network. To guarantee the fairness, we define the fairness as that the cell throughput of WiFi network should be higher than the pre-defined threshold. It is noted that the fairness criterion in 3GPP is a special case of proposed fairness, where the threshold is cell throughput of WiFi network under the coexistence of two WiFi networks \cite{LAA2015}. Specifically, the reward is redesigned as
	\begin{equation}
	\begin{aligned}
	&R^{t+1}=\begin{cases}
	(N_{\mathrm{w}}^{t}+N_{\mathrm{n}}^{t})*\mathrm{S_{file}}&,N_{\mathrm{w}}^{t}*\mathrm{S_{file}}\geq\mathrm{T_{wifi}}\\
	0&N_{\mathrm{w}}^{t}*\mathrm{S_{file}}<\mathrm{T_{wifi}},
	\end{cases}
	\end{aligned}
	\end{equation}
	where $\mathrm{T_{wifi}}$ is the pre-defined cell throughput threshold of the WiFi network. Here, the $\mathrm{T_{wifi}}$ is set as the cell throughput of WiFi network in the benchmark scheme with fixed ED thresholds. As the FTP-3 traffic follows Poisson arrival with fixed arrival rate, we further verify the proposed DRL algorithm with more dynamic Beta traffic \cite{NavarroOrtiz2020}.

	\begin{figure}[!h]
			\centering
			\subfigure[gNB-Cat4/UE-Cat2 LBT with FTP-3 traffic]
			{\includegraphics[scale=0.45]{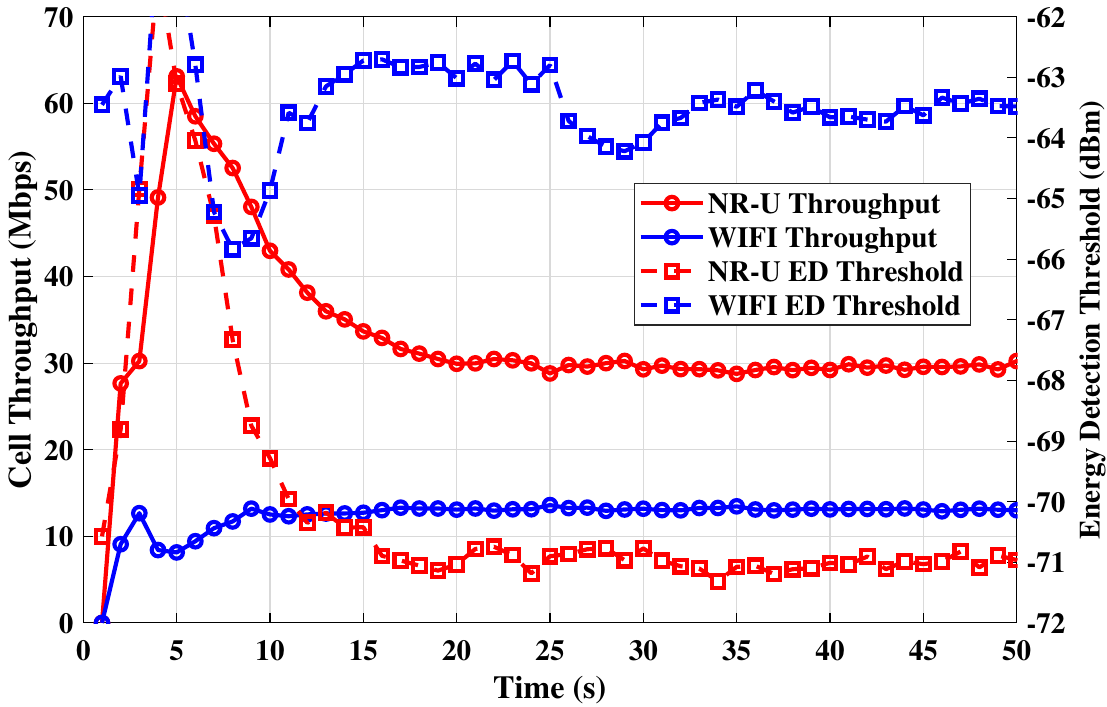}} 
			\\
			\centering
			\subfigure[gNB-Cat4/UE-Cat2 LBT with Beta traffic]
			{\includegraphics[scale=0.45]{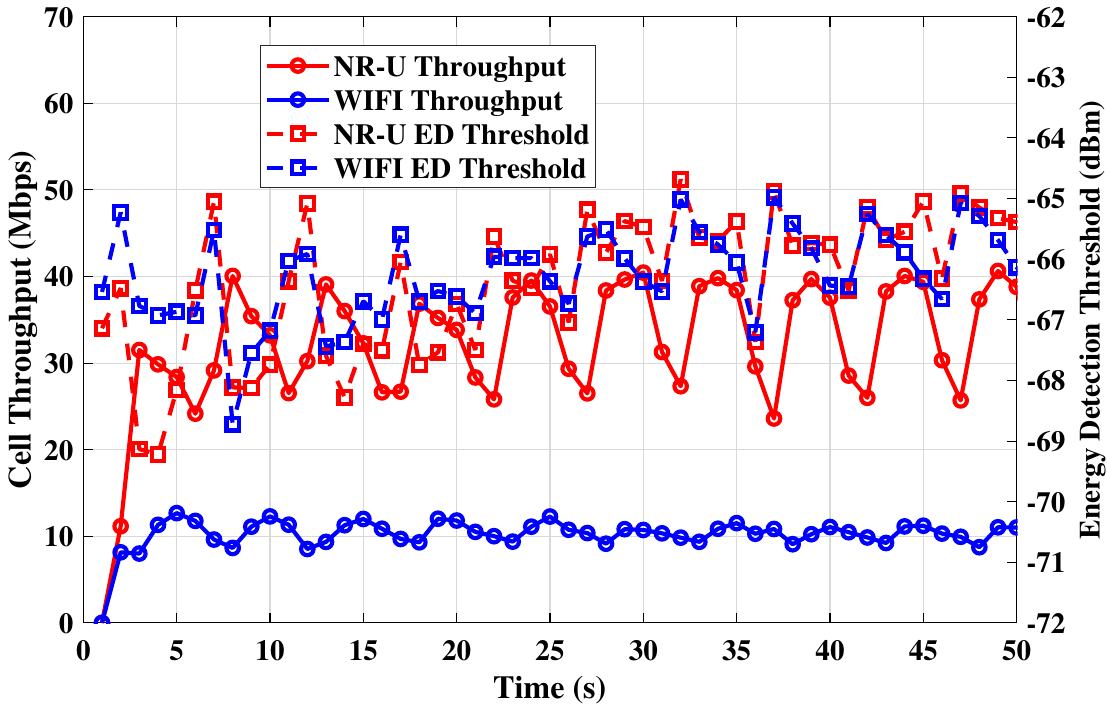}}
			\caption{Cell throughput and actions of WiFi and NR-U networks during testing phase.}
			\label{Rate_eMBB_fairness}
	\end{figure}
	
	As shown in the Fig.~\ref{Rate_eMBB_fairness}, we test the performance of converged trained model, where the average cell throughput and ED thresholds of NR-U and WiFi networks are characterized, respectively. We can see that the networks reach the stable state around 40 seconds under FTP-3 traffic, but the networks dynamically change over time under Beta traffic. Under FTP-3 traffic in the Fig.~\ref{Rate_eMBB_fairness}(a), we can see 
	that the cell throughput of NR-U network stabilizes around 31 Mbps, which is lower compared to the 58 Mbps in Fig.~\ref{Testing}. However, the cell throughput of WiFi network stabilizes around 13 Mbps, which is higher than the 6 Mbps in Fig.~\ref{Testing}. This can be explained that the NR-U network adopts -71 dBm ED threshold, but WiFi network adopts a higher ED threshold around -64 dBm. Under Beta traffic in the Fig.~\ref{Rate_eMBB_fairness}(b), we can observe that the ED thresholds of NR-U and WiFi networks have a similar dynamic range from -65 dBm to -67 dBm. However, the cell throughput dynamic range of NR-U network is much larger than that of WiFi network. This is because the fixed -82 dBm CS threshold limits the transmission of WiFi network.

	\begin{figure}[!h]
		\centerline{\includegraphics[scale=0.4]{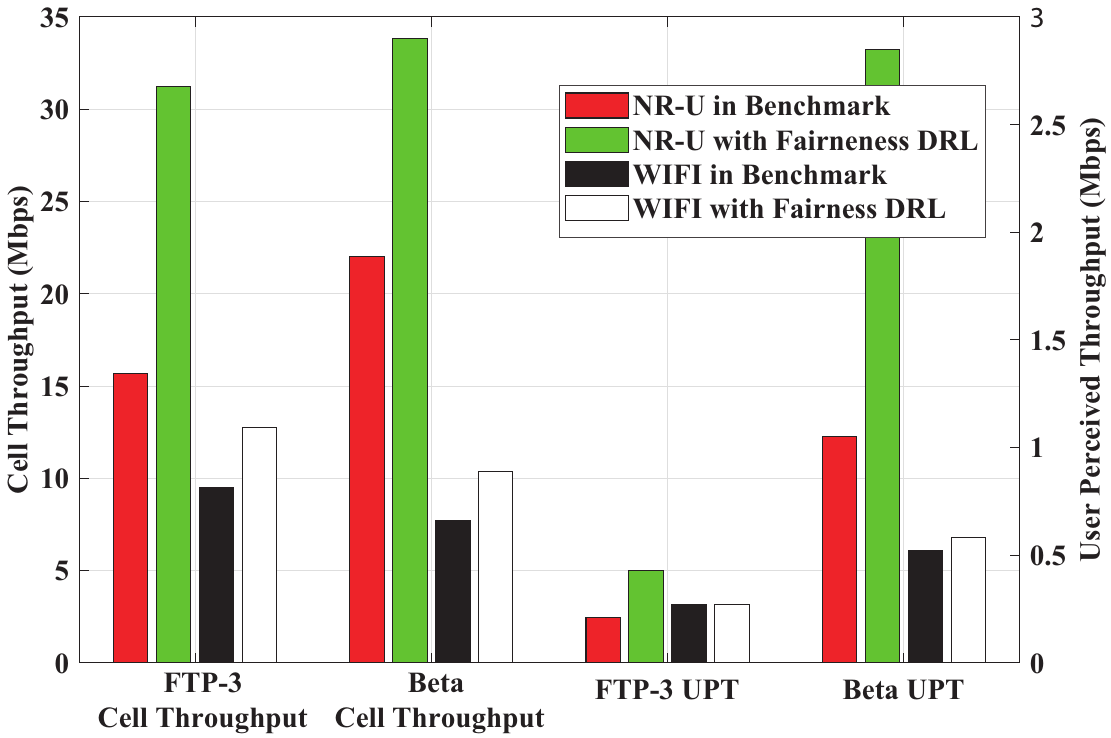}}
		\caption{Average cell throughput and user perceived throughput during testing phase in gNB-Cat4/UE-Cat2 scheme.}
		\label{fairness_UPT}
	\end{figure}
	
	Fig.~\ref{fairness_UPT} plots the average cell throughput and UPT of the proposed DRL algorithm with a re-designed reward function and benchmark scheme with fixed ED thresholds. Overall, we can see that the cell throughput of WiFi network is not sacrificed while improving the system throughput. Under FTP-3 traffic, we can see that the cell throughput of NR-U network increases from 16 Mbps to 31 Mbps with around 100\% performance gain, and the cell throughput of WiFi network increases from 9.5 Mbps to 13 Mbps with performance gain over 35\%. The system throughput increases from 25.5 Mbps to 44 Mbps, where the performance gain is over 70\%. The performance gain of the system throughput decreases compared to the algorithm without considering the fairness due to the lower ED threshold in NR-U network, which indicates the effectiveness of the re-designed reward. We can observe that the UPT of NR-U network increases from 0.21 Mbps to around 0.43 Mbps with performance gain over 100\%, which shows the re-designed reward has little influence on the UPT of NR-U network. However, UPT of WiFi network remains around 0.27 Mbps, which is a large improvement compared to the result without fairness. Under Beta traffic, we can see that the cell throughput performance gain of NR-U and WiFi networks are around 50\%, and 30\%, respectively, which leads to the system throughput gain around 48\%.

	\section{Conclusion and future work}
	In this paper, we developed a novel centralized DRL framework and a federated DRL framework, respectively, to optimize the ED thresholds configuration for maximizing the uplink system throughput in UCBC under the coexistence of heterogeneous NR-U and WiFi networks. We first developed a centralized double Deep Q-Network (DDQN) algorithm for the dynamic ED thresholds configuration with an agent deployed at the central server, where two different uplink transmission schemes (i.e., gNB-Cat4/UE-Cat2 and gNB-Cat4/UE-Cat4) are considered. To protect the data privacy, we further developed a federated DDQN algorithm to dynamically optimize the ED thresholds for NR-U and WiFi networks, where two independent agents are deployed in the heterogeneous NR-U and WiFi networks and only exchange the model parameters with the central server, respectively. Furthermore, we re-designed
	the reward function to guide the agent to guarantee the cell throughput of WiFi network while improving the uplink system throughput, where the agent gets punished when the cell throughput of WiFi network is below the pre-defined threshold. Finally, we introduced a user-centric performance metric called UPT to evaluate the file transmission throughput from the perspective of users.
	
	Our results demonstrated that our proposed DRL-based ED thresholds configuration approaches significantly outperform the benchmark scheme in terms of both uplink system throughput and UPT. Our numerical results shed light on that the NR-U network has inherent advantages over the WiFi network, including the data rate, MCOT length, scheduling policy, etc, which makes the fairness among heterogeneous networks to be an important problem to study. 
	
%	Furthermore, the coexistence of WiFi networks causes more complex interference due to the lack of scheduling, where the fully-decentralized DRL could be a promising solution.
	
	\ifCLASSOPTIONcaptionsoff
	\newpage
	\fi

	\bibliographystyle{IEEEtran}
	
	\bibliography{IEEEabrv,work2}
\end{document}